\documentclass[11pt]{article}

\usepackage{arxiv}

\usepackage[T1]{fontenc}    
\usepackage{color}
\usepackage{hyperref}       
\usepackage{url}            
\usepackage{booktabs}       
\usepackage{amsfonts}       
\usepackage{amsmath}        
\usepackage{nicefrac}       
\usepackage{microtype}      
\usepackage{graphicx}
\usepackage{lipsum}
\usepackage{lmodern}
\usepackage{subfigure}

\def\cD{{\cal D}}
\def\cE{{\cal E}}
\def\cF{{\cal F}}

\def\cI{{\cal I}}

\def\cN{{\cal N}}

\def\cR{{\cal R}}

\def\vx{\mathbf{x}}
\def\vr{\mathbf{r}}
\def\vv{\mathbf{v}}
\def\vR{\mathbf{R}}
\def\vd{\mathbf{d}}
\def\RR{\mathbb{R}}
\newcommand{\mc}[1]{\mathcal{#1}}
\newcommand{\abs}[1]{\lvert#1\rvert}

\newcommand{\ud}{\,\mathrm{d}}

\title{Deep Density: circumventing the Kohn-Sham equations via symmetry preserving neural networks}

\author{Leonardo Zepeda-N\'u\~nez \\
        Department of Mathematics \\ 
        University of Wisconsin-Madison\\ 
        Madison, WI 53706\\
        \texttt{zepedanunez@wisc.edu}\\
\And        
        Yixiao Chen \\
        Program in Applied and Computational Mathematics \\
        Princeton University \\
        Princeton, NJ 08544\\
        \texttt{yixiaoc@princeton.edu}\\
 \And
        Jiefu Zhang \\
        Department of Mathematics\\
        University of California, Berkeley\\
        Berkeley, CA 94720 \\
        \texttt{jiefuzhang@berkeley.edu}
 \And
        Weile Jia  \\
        Department of Mathematics\\
        University of California, Berkeley\\
        Berkeley, CA 94720\\
        \texttt{jiaweile@berkeley.edu}\\
 \And
        Linfeng Zhang \\
        Program in Applied and Computational Mathematics \\
        Princeton University \\
        Princeton, NJ 08544 \\
        \texttt{linfengz@princeton.edu}\\
\And
        Lin Lin \\
        Department of Mathematics\\ 
        University of California, Berkeley\\
        Computational Research Division, \\
        Lawrence Berkeley National Laboratory, \\
        Berkeley, CA 94720, USA\\
        \texttt{linlin@math.berkeley.edu}
}

\begin{document}

\maketitle

\begin{abstract}
The recently developed Deep Potential [Phys. Rev. Lett. 120, 143001, 2018] is a powerful method to represent general inter-atomic potentials using deep neural networks. The success of Deep Potential rests on the proper treatment of locality and symmetry properties of each component of the network. In this paper, we leverage its network structure to effectively represent the mapping from the atomic configuration to the electron density in Kohn-Sham density function theory (KS-DFT). 
By directly targeting at the self-consistent electron density, we demonstrate that the adapted network architecture, called the Deep Density, can effectively represent the electron density as the linear combination of contributions from many local clusters. The network is constructed to satisfy the translation, rotation, and permutation symmetries, and is designed to be transferable to different system sizes. 
We demonstrate that using a relatively small number of training snapshots, Deep Density achieves excellent performance for one-dimensional insulating and metallic systems, as well as systems with mixed insulating and metallic characters. We also demonstrate its performance for real three-dimensional systems, including small organic molecules, as well as extended systems such as water (up to $512$ molecules) and aluminum (up to $256$ atoms).
\end{abstract}

 
\section{Introduction}
Kohn-Sham density function theory (KS-DFT)~\cite{KohnSham1965} is the most widely-used electronic structure theory because the electron density completely determines the ground state~\cite{HohenbergKohn1964} and the thermal~\cite{Mermin1965} properties of a quantum many-body system. The goal of KS-DFT is to obtain a mapping of the atomic configuration to the electron density, denoted by $\varrho(\vr, \{\vR_I\}_{I = 1}^{N_a})$. Here $\vR_I$ is the position of the $I$-th nuclei, $\vr$ is the electronic position, and $N_{a}$ is the number of atoms. Notwithstanding its enormous success, KS-DFT still suffers from two significant challenges. The first challenge is its computational cost; the cost of KS-DFT calculations typically scales cubically with respect to the system size, and hence the calculations can be expensive for large systems. Despite the availability and development of linear scaling methods~\cite{Goedecker1999,BowlerMiyazaki2012}, they are only applicable to treating insulating systems with relatively large energy gaps. 
The second, and more fundamental, challenge is its accuracy, which cannot be systematically improved due to the limitation of the available exchange-correlation functionals~\cite{Martin2008}.

The recent surge in applications of machine learning methods to scientific computing problems provides an alternative route for revisiting both problems. If one can find a more effective mapping for $\varrho$ using, e.g., a neural network, we can bypass the solution of the Kohn-Sham equations, and directly obtain the self-consistent electron density for a given atomic configuration. This may drastically reduce the computational time, particularly for large systems. We may further use such a network to encode the electron density obtained from theories that are more accurate than standard KS-DFT, such as the density matrix renormalization group~\cite{White1992} or the coupled-cluster theory~\cite{BartlettMusial2007}. Based on such considerations, the representation of the electron density has received much attention in the past few years~\cite{BrockherdeVogtLiEtAl2017,GrisafiFabrizioMeyerEtAl2019,RyczkoStrubbeTamblyn2019,BogojeskiBrockherdeVogt-MarantoEtAl2018,FabrizioMeyerCeriottiEtAl2019}. 
These approaches are typically based on carefully hand-crafted descriptors that encode the atomic configuration, 
a projection of $\varrho$ onto a basis,
and a machine learning algorithm mapping the descriptors to the coefficients of the projection. 
For example, in~\cite{BrockherdeVogtLiEtAl2017} the authors adopted a descriptor using a fictitious potential centered at each nuclei, which is then mapped to a Fourier basis using a ridge kernel regression algorithm. 

The problem of using machine learning methods to represent the electron density is closely related to the problem of finding the interatomic potential and corresponding force field for molecular dynamics simulation~\cite{behler2007generalized, bartok2010gaussian,rupp2012fast,montavon2013machine,chmiela2017machine,schutt2017schnet,smith2017ani,han2017deep}. In particular, the recently developed Deep Potential scheme~\cite{ZhangHanWangEtAl2018,ZhangHanWangEtAl2018a} has been successful in describing with high fidelity various finite-size and extended systems, including organic molecules, metals, semiconductors, and insulators. 

In this work, we leverage the construction of the Deep Potential to build a neural network representation for $\varrho$. 
In contrast with related approaches, we learn the descriptor on the fly, and instead of learning the mapping to the coefficients of a basis, we evaluate the total electron density directly in a point-wise manner. The total electron density is decomposed into a linear combination of $N_{a}$ components, with the $I$-th component describing the contribution to the electron density from the $I$-th atom and its neighbors.
Each component is constructed to locally satisfy the necessary translation, rotation, and permutation symmetries. The number of atoms involved in each component does not scale with respect to the global system size, and hence the cost for evaluating $\varrho$ scales linearly with respect to the system size. 

By targeting the self-consistent electron density, we demonstrate that Deep Density can take advantage of the screening effect to effectively represent $\varrho$ for both insulating and metallic systems, and thus can bypass the solution of the nonlinear Kohn-Sham equations. The resulting algorithm is not tied to a given discretization scheme nor a particular choice of basis, and it can be transferred to systems of larger sizes. The algorithm can also be implemented in an embarrassingly parallel fashion: one can evaluate different points of the density given by different configuration completely independently. The total cost of evaluating the density scales linearly with the number of atoms in the system. 

We demonstrate that Deep Density can accurately predict the electron density and exhibits excellent transferability properties for one-dimensional model systems of insulating, metallic, as well as mixed insulating-metallic characters. We also report the performance of our model for three-dimensional real systems, including small molecules  (C$_2$H$_6$, C$_4$H$_{10}$), as well as condensed matter systems, including water (up to $512$ water molecules), and aluminum (up to $256$ aluminum atoms).

The rest of the paper is organized as follows. In Section~\ref{section:preliminaries} we provide the framework of KS-DFT, and the map we seek to approximate. In Section~\ref{section:architecture} we provide the architecture for the neural network. In particular, we provide a succinct review of the techniques in \cite{zhang2019active} and we explain how to implement the physical ansatz and the symmetry requirements. 
In Section~\ref{section:numerical_examples} we provide the numerical examples showcasing the accuracy and transferability of the algorithm. In particular, we provide the electron density for $1$D models (Section~\ref{sec:onedim}) and realistic $3$D systems (Section~\ref{sec:threedim}) and we compare them against the electron density produced by classical KS-DFT computations. 
We show that it is possible to train these models to single precision in $1$D and within three digits in $3$D, and that the accuracy is maintained even for test systems an order of magnitude larger. 
In Section \ref{sec:conclude} we provide several comments about the current work and we point to several future directions of research. Additional details for the numerical experiments are given in the appendices.

\section{Preliminaries} \label{section:preliminaries}

For a system with $N_e$ electrons at a given atomic configuration $\{\vR_I \}_{I = 1}^{N_a}$ in a $d$-dimensional space (i.e. $\vr,\vR_I\in\RR^d$), KS-DFT solves the following nonlinear eigenvalue problem (spin omitted for simplicity)
\begin{align}
H[\rho;\{\vR_I \}] \psi_{i}& =\varepsilon_{i} \psi_{i}, \quad i = 1, ..., N_e, \\
\int \psi_{i}^{*}(\vr) \psi_{j}(\vr) \ud \vr & =\delta_{i j}, \quad \rho(\mathbf{r}) =\sum_{i=1}^{N_e}\left|\psi_{i}(\vr)\right|^{2}. 
\end{align}
The Kohn-Sham Hamiltonian is
\begin{equation}
  H[\rho;\{\vR_I \}]:=-\frac{1}{2} \Delta_{\vr}+V_{\mathrm{ion}}(\vr;\{\vR_I\})+V_{\mathrm{hxc}}[\vr;\rho].
\end{equation}
Here $V_{\mathrm{ion}}$ characterizes the interaction between electrons and nuclei, and it does not depend on the electron density. $V_{\mathrm{hxc}}$ is called the Hartree-exchange-correlation potential, which includes the mean-field electron-electron interaction, as well as the contribution from the exchange-correlation energy. The eigenvalues $\{\varepsilon_i\}_{i =1}^{N_e}$ are real and ordered non-decreasingly, so $\{\psi_{i}\}_{i=1}^{N_e}$ correspond to the eigenfunctions with lowest $N_e$ eigenvalues. Due to the $\rho$-dependence of $V_{\mathrm{hxc}}$, the Kohn-Sham equations needs to be solved self-consistently till convergence. We refer to \cite{Martin2008,LinLuYing2019} for more details of numerical solutions of KS-DFT. For the purpose of this paper, we are interested in learning the mapping 
\begin{equation}
  \varrho: \{\vr\} \cup \{\vR_I\}_{I = 1}^{N_a} \mapsto \rho(\vr),
  \label{eqn:mappingrho}
\end{equation}
which is also denoted as $\varrho(\vr,\{\vR_I\}_{I = 1}^{N_a})=\rho(\vr)$. 

In KS-DFT, we need to distinguish between the external potential $V_{\mathrm{ion}}(\vr;\{\vR_I\})$, and the effective potential
\begin{equation}
  V_{\mathrm{eff}}(\vr)=V_{\mathrm{ion}}(\vr;\{\vR_I\})+V_{\mathrm{hxc}}[\vr;\rho].
  \label{}
\end{equation}

We refer to the mapping from $V_{\mathrm{eff}}$ to $\rho$ as the linearized Kohn-Sham map. The evaluation of the Kohn-Sham equations requires (partially) diagonalizing the Hamiltonian $H_{\mathrm{eff}}:=-\frac12 \Delta_{\vr}+V_{\mathrm{eff}}(\vr)$, after proper discretization. Correspondingly we refer to the mapping from $V_{\mathrm{ion}}$ to $\rho$ as the self-consistent Kohn-Sham map, of which the evaluation requires solving the Kohn-Sham equations self-consistently. 

For insulating systems with a positive energy gap (i.e. $\varepsilon_{N_{e}+1}-\varepsilon_{N_e}>0$), it is known that the dependence of electron density at position $\vr$ with respect to the potential at position $\vr'$ decays exponentially with respect to $\abs{\vr-\vr'}$. This is often referred to as the ``nearsightedness'' principle of electrons~\cite{Kohn1996,ProdanKohn2005}. More specifically, the Fr\'{e}chet derivative
\begin{equation}
  \chi_0(\vr,\vr') = \frac{\delta \rho(\vr)}{\delta V_{\mathrm{eff}}(\vr')},
  \label{}
\end{equation}
which is also called the irreducible polarizability operator in physics literature, satisfies the decay property $\chi_{0}(\vr,\vr')\sim e^{-c \abs{\vr-\vr'}}$ for some constant $c>0$. The nearsightedness principle allows the design of linear scaling methods~\cite{Goedecker1999,BowlerMiyazaki2012}, which effectively truncate the global domain into many small domains to solve KS-DFT. 

For metallic systems with a zero or very small energy gap, $\chi_{0}(\vr,\vr')$ only decays algebraically as $\abs{\vr-\vr'}\to \infty$. Hence the nearsightedness principle does not hold anymore, and linear scaling methods become either very expensive (with a large truncation radius) or inaccurate (with a small truncation radius). On the other hand, even for metallic systems, the reducible polarizability operator, defined as 
\begin{equation}
  \chi(\vr,\vr') = \frac{\delta \rho(\vr)}{\delta V_{\mathrm{ion}}(\vr')},
  \label{}
\end{equation}
can be much more localized compared to $\chi_{0}$. This is known as the screening effect~\cite{Eguiluz1985,GonzeLee1997}.  

\begin{figure}[htbp]
\begin{center}
{%
\subfigure[Perturbation $\delta V_{\mathrm{ion}}$]{%
\label{fig:Vext}
\includegraphics[width=0.35\textwidth]{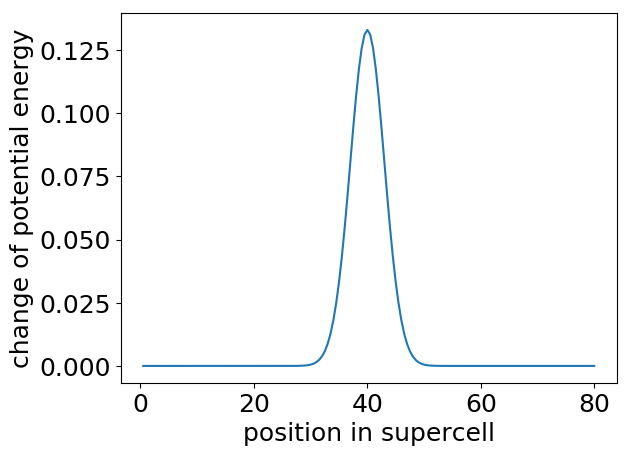}
} 
\qquad
\subfigure[Comparison of $\chi \delta V_{\text{ion}}$ and $\chi_0\delta V_{\text{ion}}$ ]{%
\label{fig:chiVschi0}
\includegraphics[width=0.35\textwidth]{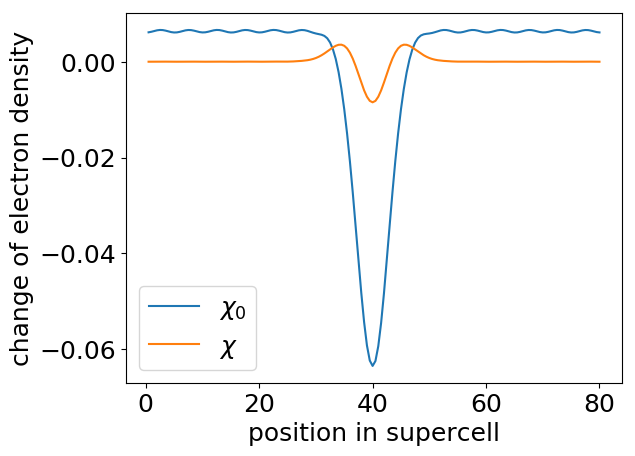}
}
}
\end{center}
\caption{Illustration of the screening effect for a one-dimensional model metallic system.}
\label{fig:1Dscreen}
\end{figure}

To illustrate the screening effect, we consider a one-dimensional periodic metallic system with $8$ atoms. The details of the setup will be discussed in Section~\ref{sec:onedim} and Appendix \ref{app:numer1D}.  
We introduce a localized and exponentially decaying perturbation of the potential $\delta V_{\mathrm{ion}}$ as
\begin{equation}
    \delta V_{\mathrm{ion}}(r) = \frac{1}{\sqrt{2\pi \sigma_0^2}}\exp{\left(-\frac{1}{2\sigma_0^2}(r-\mu_0)^2\right)},
\end{equation}
where we choose $\sigma_0=3.0$, and $\mu_0=40$ is the center of the supercell. Fig.~\ref{fig:Vext} displays the profile of $\delta V_{\mathrm{ion}}$. Fig.~\ref{fig:chiVschi0} shows that for a metallic system, the induced electron density obtained from the linearized Kohn-Sham map, which can be approximately computed as $\chi_{0}\delta V_{\text{ion}}$, has a large magnitude, and a delocalized and oscillatory tail. On the other hand, due to the screening effect, the magnitude of the induced electron density obtained from the self-consistent Kohn-Sham map, which can be approximately computed as $\chi \delta V_{\text{ion}}$, has a much smaller magnitude. Its tail is much shorter and smoother, and the support of  $\chi \delta V_{\text{ion}}$ is localized around that of $\delta V_{\text{ion}}$. The screening effect indicates that Deep Density should be able to leverage such an additional level of localization which is not present in standard linear scaling solvers, and a relatively small truncation radius may be possible even for metallic systems (see the architecture in Section \ref{section:architecture}).

\section{Network architecture} \label{section:architecture}

\subsection{Locality}
 
Standard methods for solving KS-DFT can only be used to evaluate the linearized Kohn-Sham map, which is then used to obtain the self-consistent electron density via self-consistent field iterations. In contrast, our goal is to {\it directly} learn the self-consistent electron density, or the mapping $\varrho(\vr,\{\vR_I\}_{I = 1}^{N_a})$, which already takes into account the screening effect. This allows us to construct neural networks that can be decomposed into the linear combination of local components for both insulating and metallic systems.  

\begin{figure}
    \centering
    \includegraphics[width=0.5\textwidth, trim={0mm 0mm 0mm 0mm},clip]{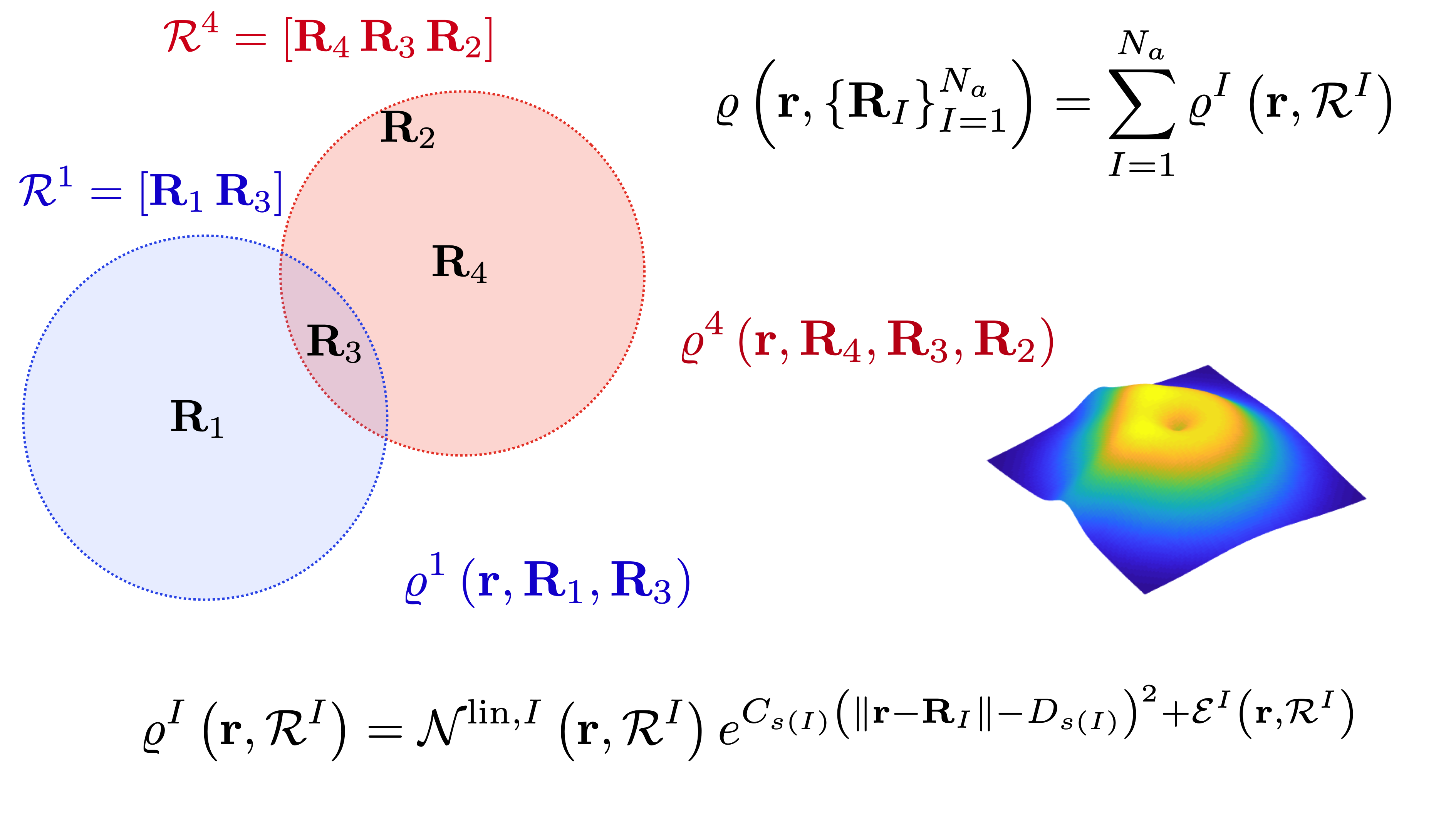}
    \caption{Illustration of the local interaction lists and local electron densities.}
    \label{fig:interaction_sketch}
\end{figure}

As illustrated in Fig.~\ref{fig:interaction_sketch}, we partition the electron density as
\begin{equation} \label{eq:sum_rho}
  \varrho(\vr, \{\vR_I\}_{I = 1}^{N_a}) = \sum_{I=1}^{N_a} \varrho^I(\vr,\cR^I).
\end{equation}
Here $\varrho^I$ characterizes the contribution to the electron density at $\vr$ from the neighborhood of the $I$-th atom. The locality is imposed by building an interaction list $\mathring{\cI}_{R_c}(I)$, defined as the set of indices $J$ such that $\abs{\vR_J-\vR_I}< R_c$.  Following this notation, $\cR^I$ denotes the set of atomic positions
\[
\cR^I = \{\vR_J, J \in \mathring{\cI}_{R_c}(I)\}.
\]
We denote by $s(I)$ the species index (i.e. the atomic number) of the $I$-th atom. We may also treat the electron as a special ``atom'' equipped with index $J=0$. For simplicity we define $s(0)=0,\vR_{0}=\vr$, and then define the extended interaction list as the index set
\begin{equation}
  \cI_{R_c}(I)=\{0\}\cup \mathring{\cI}_{R_c}(I).
  \label{eq:index}
\end{equation}
In other words, the electron (formally) belongs to  $\cI_{R_c}(I)$ for every atom $I$.

The density $\varrho^I$ can be constructed as
\begin{equation} \label{eq:ansatz}
  \varrho^I (\vr, \cR^I) = \cN^{\texttt{lin},I} (\vr,\cR^I) e^{C_{s(I)}\left( \abs{\vr - \vR_I} - D_{s(I)}\right)^2 + \cE^{I} (\vr,\cR_I)}.
\end{equation}
We assume $\cN^{\texttt{lin},I}$ takes the form 
\begin{equation}
  \cN^{\texttt{lin},I}_{s(I)} (\vr,\cR^I) = \cN^{I} (\vr,\cR_I)  + A_{s(I)}\abs{\vr - \vR_I} +B_{s(I)}.
\end{equation}
The constants $A, B, C, D$ are trainable parameters and only depend on the species of the $I$-th atom.

The ansatz is built such that the electron density decays exponentially with the distance of the electron to the center of the cluster, i.e., the $I$-th atom. The rate of decay is given by $A$, and $B$ accounts for the possibility that the bulk of the electron density is not centered at the atom $I$. In addition, due to possible issues with the pseudopotential in the KS-DFT computation, the electron density can be very small in the neighborhood of the nuclei. In order to fit this the nonlinear correction, $\cE^{I} (\vr,\cR_I)$ would need to have a large negative value. Thus, in order to bypass this issue we multiply the exponential with a linear term to capture this behavior. Moreover, to capture abrupt changes on the density we add $\cN^I$, a non-linear correction, to this term. 

In a nutshell, the exponential term mimics the behavior of the local density far from the center of the cluster, whereas $\cN^{\texttt{lin},I}$ captures the behavior close to center of atom $I$. $\cN^I,\cE^I$ are neural networks to be introduced later. Clearly we may absorb the constants $A, B, C, D$ into the neural networks as well. In practice, we found that separating out such terms can reduce the training time and the test error of the network. In addition, numerically we observed that the derivative associated to the constant may widely differ from the rest of the neural network, introducing them explicitly allows us to re-scale the gradient if necessary, thus accelerating the optimization. 

%

\subsection{Symmetry}
When one considers modeling the physics at different scales, it is often crucial to preserve symmetry properties. 
Additionally, from a learning perspective, symmetries can also significantly improve the efficiency of training in terms of the required number of parameters, the number of training steps, and the amount of training data.
In this regard, as a mapping from electronic and atomic positions to the   value of the electron density, $\varrho(\vr,\{\vR_I\}_{I = 1}^{N_a})$ should be invariant under permutation of  indices of identical atoms, and under a collective translation or rotation of the electron and atomic positions.
As mentioned in the introduction, the problem can also be considered as finding a map from atomic positions to a electron density field represented by electron positions or other basis sets.
One disadvantage is that the dimension of the output may be large: the output consists of values of the density on a list of grid points, or the coefficients corresponding to a basis set. When the output is a list of grid points, it can also be difficult to satisfy symmetry requirements. Here we follow the first approach, and we find  it is much easier to define a neural network representation with a single output.
Finally, to combine the locality and the symmetry requirements, we guarantee that the values of $\cN^{I}$ and $\cE^{I}$ in Deep Density are invariant with respect to the following symmetry operations:
\begin{enumerate}
  \item Permutation: relabeling of indices of the same atom species in the index set $\cI_{R_c}(I)$.
  \item Translation: uniformly shifting $\vR_I$ and the positions of all particles in $\cI_{R_c}(I)$ by a vector.
  \item Rotation: rotating all particles in $\cI_{R_c}(I)$ around $\vR_I$.
\end{enumerate}
In order to reduce the number of parameters, all neural networks involved should share the same set of parameters if the species of the particles involved are the same.


We now illustrate the treatment of $\cN^{I}$; the procedure for $\cE^I$ is analogous. In order to satisfy permutation symmetry, we adopt a variant of the representation in~\cite{ZaheerKotturRavanbakhshEtAl2017}, which states that any permutation invariant function $f:\RR^N\to\RR$ can be represented in the form 
\begin{equation}
  f(\vx)=f(x_1,\ldots,x_N)=\mc{F}\left(\sum_{i=1}^N h(x_i)\right),
  \label{}
\end{equation}
where $h:\RR\to\RR^M$ maps each coordinate $x_i$ into an $M$-dimensional  feature space, and $\mc{F}:\RR^M\to\RR$ combines the features to obtain the value $f$. Both $\mc{F}$ and $h$ can are represented by neural networks in what follows.

In this work, we follow the construction in~\cite{ZhangHanWangEtAl2018a}, and decompose $\cN^I$ as 
\begin{equation} \label{eq:decomp_descriptor}
    \cN^I(\vr, \cR^I) = \cF_{s(I)} \circ \cD^I (\vr, \cR^I).
\end{equation}
Here $\cF_{s(I)}: \RR^{M\times M} \rightarrow \mathbb{R}$ is called a fitting network that only depends on the species index $s(I)$. $\cD^I(\vr, \cR^I)$ is called a descriptor network, and $\cD^I(\vr, \cR^I)\in \RR^{M\times M}$ is a matrix of the following form
\begin{align}
  \cD^I (\vr, \cR^I) = \left(\sum_{J\in \cI_{R_c}(I)} h_{s(I),s(J)}(\vR_J-\vR_I)\right)^{\top} \left(\sum_{J\in \cI_{R_c}(I)} h_{s(I),s(J)}(\vR_J-\vR_I)\right).
  \label{eqn:dnetwork}
\end{align}
Here $h_{s(I),s(J)}:\RR^d\to\RR^{\tilde{d}\times M}$ is a feature mapping, which only depends on the species of the particle $I$ and $J$ (recall that the electron is labeled with $J=0$ with species index $0$). Then $\cD^{I}\in\RR^{M\times M}$ is a symmetric matrix that naturally satisfies the permutation and translation symmetries.
As will be demonstrated below, it satisfies the rotation symmetry as well.

Now we demonstrate the construction of the mapping $h_{s(I),s(J)}$. Define $R_{JI}=\abs{\vR_J-\vR_I}$, and 
we would like to require $h$ to depend smoothly on $\vR_J$ moves in and out of the index set ${\cI}_{R_c}(I)$. In other words, $h$ should continuously vanish as $R_{JI}$ approaches $R_c$. We may introduce a cutoff function
\begin{equation} \label{eq:smooth_cutoff}
  \phi\left(R\right)=\left\{
  \begin{array}{ll}{
    \frac{1}{R+\delta},} & {0\le R \le R_{c s}} \\ {\frac{1}{R+\delta}\left\{\frac{1}{2} \cos \left[\pi \frac{\left(R-R_{c s}\right)}{\left(R_{c}-R_{c s}\right)}\right]+\frac{1}{2}\right\},} & {R_{c s}<R<R_{c}} 
    \\ {0,} & {R\ge R_{c}}\end{array}\right.
\end{equation}
where  $0< R_{c s} < R_{c}$. Note that $\phi\in C^1(\mathbb{R}^{+}\cup\{0\})$ for any $\delta>0$. 

Then we may define the generalized coordinate as
\begin{equation}
  \vd^I_J = \begin{bmatrix}
  \phi(R_{JI}) \\
  \frac{\phi(R_{JI})}{R_{JI}}(\vR_{J}-\vR_I)
\end{bmatrix}\in \RR^{d+1}, \quad J \in \mathring{\cI}_{R_c}(I).
\label{eq:dIJ}
\end{equation}
Here $\tilde{d}:=d+1$ is the dimension of the generalized coordinate.  In principle we may use a different set of generalized coordinates for electrons. For simplicity, in this work we apply the same definition as in \eqref{eq:dIJ} to the electron,  with the same truncation radius $R_c$. 
Therefore, effectively, the electron at position $\vr$ only belongs to the extended interaction lists of its neighboring atoms.

We require $h$ to depend only on the generalized coordinates as 
\begin{equation}
  h_{s(I),s(J)}(\vR_J-\vR_I)= \vd^{I}_{J} \left[g_{s(I),s(J)}\left((\vd^{I}_{J})_{1}\right)\right]^{\top} \in \RR^{\tilde{d}\times M}. 
  \label{eqn:hnetwork}
\end{equation}
Here $g_{s(I),s(J)}:\RR\to\RR^M$ is a neural network that only depends on the first component of $\vd^I_{J}$ (\textit{i.e.} the radial information $R_{JI}$), and only depends on the species of the particles $I,J$. Combining Eq.~\eqref{eqn:dnetwork} and~\eqref{eqn:hnetwork}, we have
\begin{equation}
    \cD^I (\vr, \cR^I) = \sum_{J,J'\in \cI_{R_c}(I)} \left[g_{s(I),s(J)}\left((\vd^{I}_{J})_{1}\right)\right] \left[(\vd^{I}_{J})^{\top}(\vd^{I}_{J'})\right]
    \left[g_{s(I),s(J')}\left((\vd^{I}_{J'})_{1}\right)\right]^{\top}.
  \label{eqn:dnetwork_all}
\end{equation}
Both the radial information $(\vd^{I}_{J})_{1}$ and the inner product $(\vd^{I}_{J})^{\top}(\vd^{I}_{J'})$  satisfy the rotation symmetry. 
Therefore the descriptor $\mc{D}^I$ is invariant to permutation, rotation, and translation symmetry operations.  

\begin{figure}
    \centering
    \includegraphics[width=0.5\textwidth]{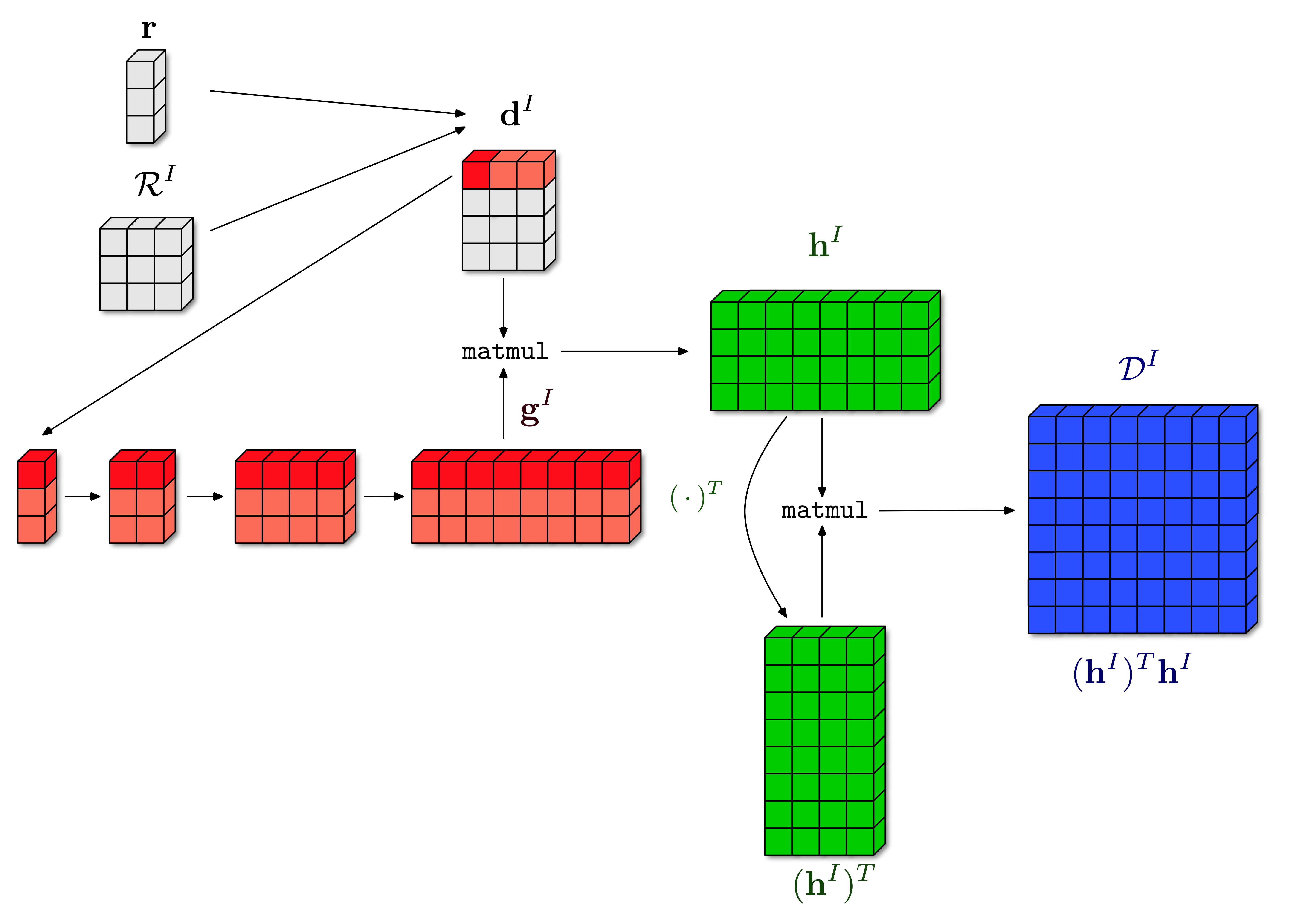}
    \caption{Schematic illustration of the computation of $\cD^I$, where $n(I) = 3$, and there is only one atomic species (besides the electron  represented in dark red).}
    \label{fig:computation_of_the descriptor}
\end{figure}

Fig.~\ref{fig:computation_of_the descriptor} provides a simplified illustration for computing $\cD^I$, where the matrix $\mathbf{g}^{I}$ encodes $\left (g_{s(I),s(J)}\right)^{\top}$ for $J \in \mathcal{I}_{R_c}(I)$ and 
\begin{align}
    \mathbf{h}^{I} &= \sum_{J\in \cI_{R_c}(I)} h_{s(I),s(J)}(\vR_J-\vR_I)= \sum_{J\in\cI_{R_c}(I)} \vd^{I}_{J} \left[g_{s(I),s(J)}\left((\vd^{I}_{J})_{1}\right)\right]^{\top}= \vd^I \mathbf{g}^{I}.
\end{align}
We use a ResNet \cite{HeZhangRenEtAl2016} architecture using dense layers to construct $\cF_{s(I)}$, while the feature mapping $g_{s(I), s(J)}$ is a dense feed-forward neural network with a few layers. 

\section{Numerical examples} \label{section:numerical_examples}

In this section we report the performance of Deep Density for one-dimensional model problems, as well as three dimensional real systems. The details of the setup as well as the choice of hyperparameters can be found in Appendix \ref{app:numer1D} and \ref{app:numer3D}, respectively. In all calculations the test error is measured in terms of the relative $\ell^1$/$\ell^2$ error, defined as: 
\begin{equation}
     \texttt{err}_{\ell^1} := \frac{ \sum_{i} \left | \varrho(\vr_{i}, \{\vR_I\}) - \varrho_{NN}(\vr_{i}, \{\vR_I\}) \right | }{ \sum_{i} \left | \varrho(\vr_{i}, \{\vR_I\})\right | },
\end{equation}
\begin{equation}
  \texttt{err}_{\ell^2} :=  \frac{\left ( \sum_{i} \left [ \varrho(\vr_{i}, \{\vR_I\}) - \varrho_{NN}(\vr_{i}, \{\vR_I\}) \right ]^2  \right)^{\frac{1}{2}}}{\left ( \sum_{i} \left [ \varrho(\vr_{i}, \{\vR_I\})\right ]^2  \right)^{\frac{1}{2}}}.
\end{equation}
Here the index $i$ is taken over all the discretization points, $\varrho$ is the electron density computed using Kohn-Sham solvers, and $ \varrho_{NN}$ is the approximation given by the neural network. In particular, the relative $\ell^1$ error approximates the following quantity
\begin{equation}
\frac{\int  \left | \varrho(\vr, \{\vR_I\}) - \varrho_{NN}(\vr, \{\vR_I\}) \right| \ud \vr }{N_e}
\label{eqn:l1_continuous}
\end{equation}
which is the same error metric used by e.g. \cite{grisafi2018transferable}. 

\subsection{One dimensional systems}\label{sec:onedim}

In this section we study three model systems in 1D with different characters: insulating, metallic, and mixed metallic-insulating systems. 
Previous study indicates that when the system is metallic or has mixed metallic-insulating characters, the self-consistent field iteration can be very difficult to converge (without a proper preconditioner) due to the small energy gaps and the associated charge sloshing behavior \cite{LinYang2013}. The details of the setup can be found in Appendix \ref{app:numer1D}.

We first consider a small supercell consisting of $8$ atoms initially separated by 10 a.u. At the beginning of the \textit{ab initio} molecular dynamics simulation, we perturb each of the 8 atoms by a uniform random number in $[-3,3]$ a.u., and then let the systems evolve for  $30000$ time steps. In order to reduce the correlation among the snapshots and the amount of training time, we down-sample the trajectory for the first $8000$ time steps by a factor $80$, and we take the resulting first 100 snapshots as the training snapshots.  The same procedure is applied to the validation snapshots for the next $400$ time steps. We then use these $100$ training snapshots and $5$ validation snapshots to train the network. 

\begin{figure}[htbp]
{%
\subfigure[ the insulating system, $\texttt{err}_{\ell^1} =$ 6.27E-04]{%
\label{fig:ins2019}
\includegraphics[width=0.3\textwidth]{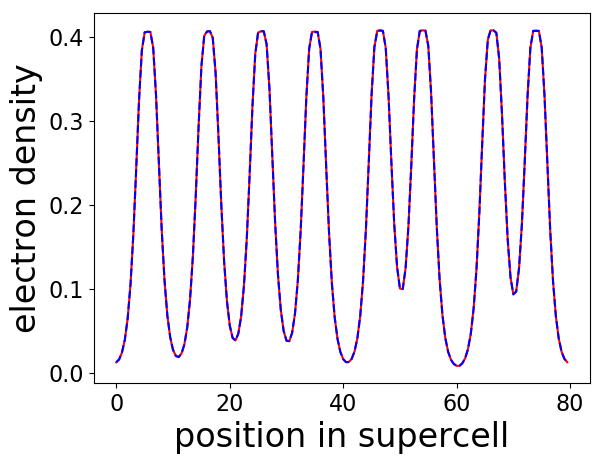}
} 
\subfigure[the metallic system, $\texttt{err}_{\ell^1} =$ 1.21E-04]{%
\label{fig:met2019}
\includegraphics[width=0.3\textwidth]{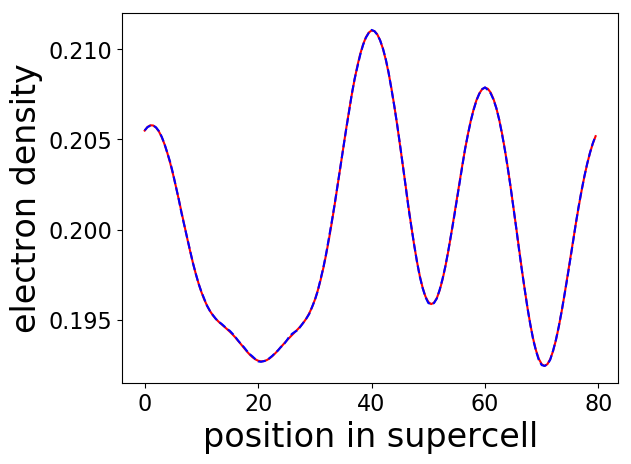}
} 
\subfigure[the mixed metallic-insulating system, $\texttt{err}_{\ell^1} =$ 5.42E-04]{%
\label{fig:halfnhalf2019}
\includegraphics[width=0.3\textwidth]{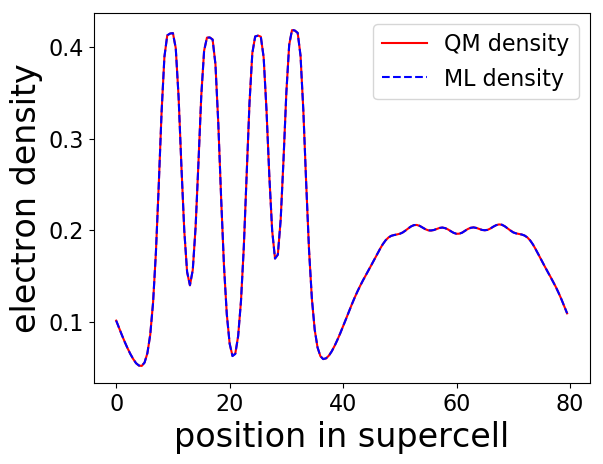}
}
\caption{Comparison of the electron density at time step $2019$.}
\label{fig:1Dinter}
}
\end{figure}

\begin{figure}[htbp]
{%
\subfigure[the insulator system, $\texttt{err}_{\ell^1} =$ 8.12E-04]{%
\label{fig:ins29000}
\includegraphics[width=0.3\textwidth]{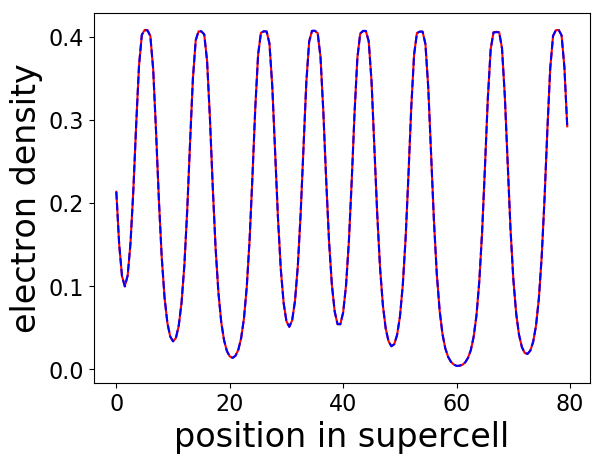}
} 
\subfigure[the metallic system, $\texttt{err}_{\ell^1} =$ 1.15E-04]{%
\label{fig:met29000}
\includegraphics[width=0.3\textwidth]{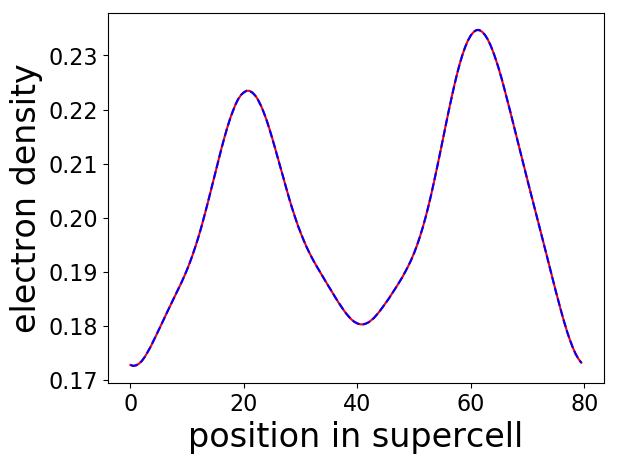}
}
\subfigure[the mixed metallic-insulating system, $\texttt{err}_{\ell^1} =$ 4.30E-03]{%
\label{fig:halfnhalf29000}
\includegraphics[width=0.3\textwidth]{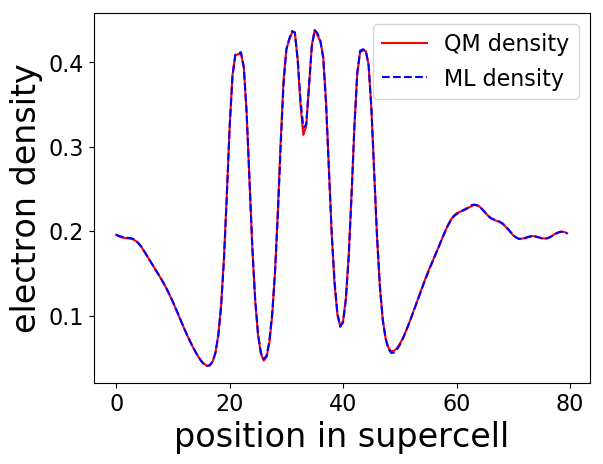}
}
\caption{Comparison of the electron density at time step $29000$.}
\label{fig:1Dtimeext}
}
\end{figure}

We test the trained model by comparing the predicted density and the density obtained from the KS-DFT calculation, using a snapshot which is part of the original training set at time step 2019 (before the down-sampling) in Fig.~\ref{fig:1Dinter}, as well as a snapshot that is far outside the training set at time step 29000 in Fig.~\ref{fig:1Dtimeext}. In both cases, the the error of the electron density is very small, and is $0.01\%\sim 0.43\%$ measured by the relative $\ell^1$ norm.

\begin{figure}[htbp]
{%
\subfigure[the insulator system, $\texttt{err}_{\ell^1} =$ 6.86E-04;]{%
\label{fig:ins32atoms}
\includegraphics[width=0.3\textwidth]{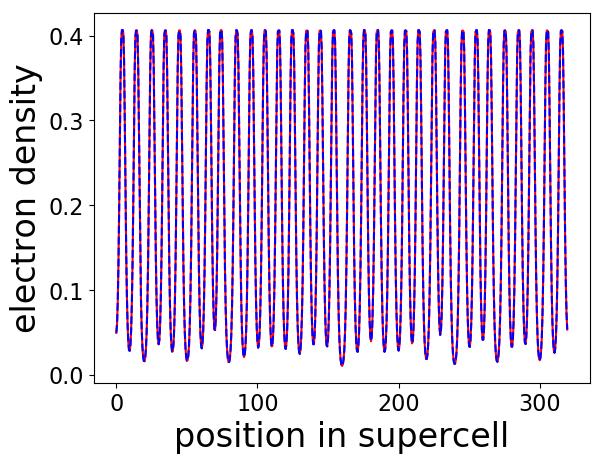}
} 
\subfigure[the metallic system, $\texttt{err}_{\ell^1} =$ 1.49E-04;]{%
\label{fig:met32atoms}
\includegraphics[width=0.3\textwidth]{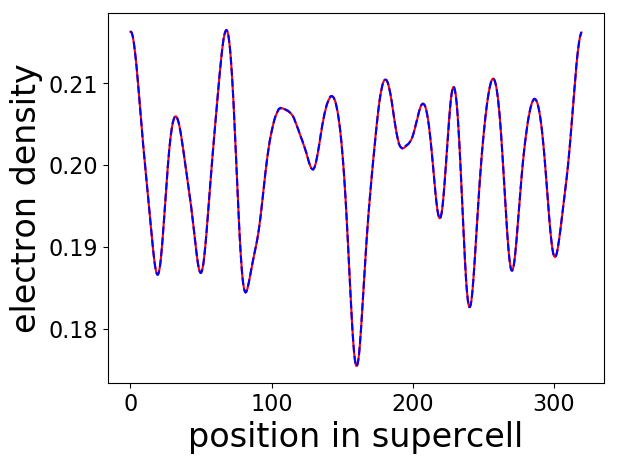}
}
\subfigure[the mixed metallic-insulating system, $\texttt{err}_{\ell^1} =$ 5.71E-03.]{%
\label{fig:halfnhalf32atoms}
\includegraphics[width=0.3\textwidth]{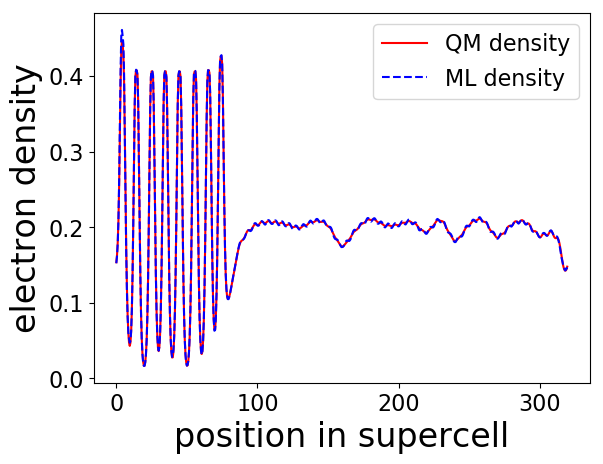}
}
\caption{Transferability of the one-dimensional model, which is trained using for a system with 8 atoms and tested on a system with 32 atoms.}
\label{fig:1Dsizeext}
}
\end{figure}

Our architecture constructs a local density $\varrho^I$ for each atom as in Eq.~\eqref{eq:sum_rho}. This enables us to use the trained model to predict the electron density of a larger system. We test the transferability of our model by loading parameters trained using the 8-atom systems into the model that predicts the electron density for the 32-atom systems. 
For the mixed metallic-insulating system, the first 8 atoms are insulator-like and the latter 24 atoms are metal-like.
Our model achieves excellent transferability, and the error is $0.01\% \sim 0.57\%$, as shown in Fig.~\ref{fig:1Dsizeext}.

\subsection{Three dimensional systems}\label{sec:threedim}

For three-dimensional molecular and condensed matter systems, we  present the following three test sets: 
\begin{itemize}

    \item organic molecules: a single  ethane molecule (C$_2$H$_6$), a single isobutane molecule and a single n-butane molecule (C$_4$H$_{10}$).    
    \item water: a set of systems composed of $32, 64, 128, 256,$ and $ 512$ water molecules in the liquid phase at T=300K.    
    \item aluminum: a set systems composed of $32, 108,$ and $256$ aluminum atoms formed initially by  $2\times2\times2$,  $3\times3\times3$, and $ 4\times4\times4$ face-center cubic (fcc) unit cells, and at temperatures 300K, 600K, and 900K, respectively.
\end{itemize}

For organic molecules, the configurations are collected from the dataset provided in \cite{cheng2019data}. We take the first 101 (uncorrelated) snapshots of ethane, n-butane and isobutane from the dataset. For each molecule, we use 100 snapshots for training and one for testing. 

For water and aluminum, the configurations are obtained using the DeePMD-kit package~\cite{wang2018kit}.
For each case, the atomic configurations are uniformly sampled from long molecular dynamics trajectories at different temperatures and different system sizes. 
For all simulations we perform N$p$T simulations at the standard pressure $p$=1 bar with a time step of 1 fs.
The potential energy models used in the simulation are obtained with the DP-GEN scheme~\cite{zhang2019active} using \textit{ab initio} data.
We take one snapshot in every 1000 time steps from the trajectory to reduce the correlation among configurations. 
The data sets are divided as follows: The training set was a randomly selected subset of $80$ snapshots from the $100$ snapshots for the smallest system. The test set was composed of the remaining $20$ snapshots for the smallest systems in addition to the snapshots of the larger ones.

For all systems, we compute the electron density for each snapshot using PWDFT (which is based on planewaves and is an independent module of the DGDFT package~\cite{HuLinYang2015a}). 
We use the Perdew-Burke-Ernzerhof (PBE) exchange-correlation functional~\cite{PerdewBurkeErnzerhof1996}, and the SG15 Optimized Norm-Conserving Vanderbilt (ONCV) pseudopotentials \cite{Hamann2013,SchlipfGygi2015}.   
Other details of the setup of  test systems and the training hyperparameters are given in Appendix \ref{app:numer3D}.  

\subsubsection{Small Molecules}
The kinetic energy cut-off is set to $30$ a.u. for both C$_2$H$_6$ and  C$_4$H$_{10}$. The total number of grid points for each system is 1,906,624. For the feature networks in Eq.~\eqref{eqn:hnetwork}, we use a dense linear network with three layers, containing $\{5,10,20\}$ nodes respectively. For each fitting network we used a ResNet \cite{he2016deep} with 3 dense layers containing 50 nodes per layer, where the skip connections are weighted by a trainable coefficient. The cutoff is the same for the 3 molecules $R_c = 4 \AA$.
We trained the network for a few times with different random seeds and picked the one with the smallest generalization error.
 
\begin{figure}
     \centering
     \includegraphics[width=\textwidth,clip]{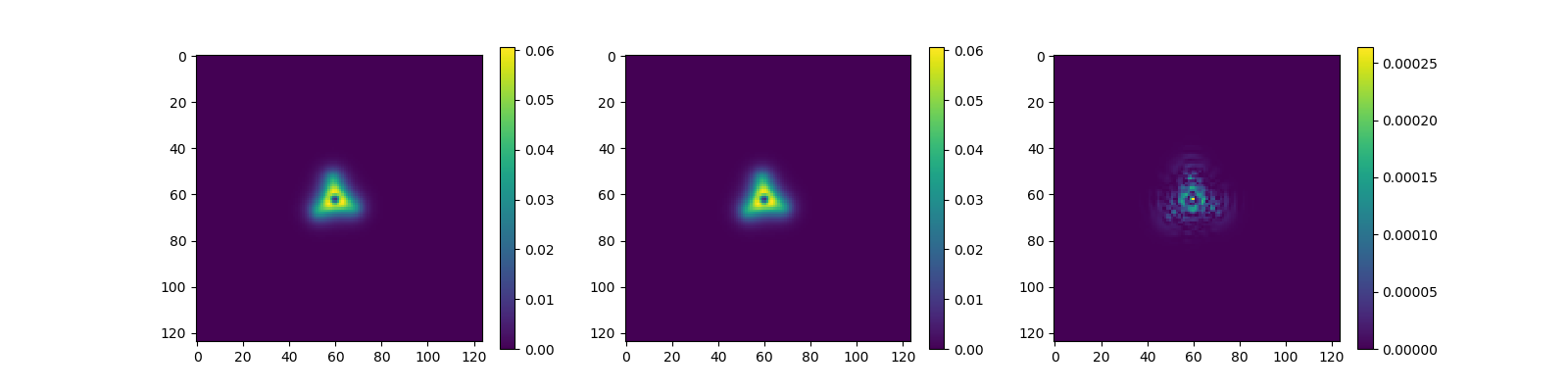}
     \includegraphics[width=\textwidth,clip]{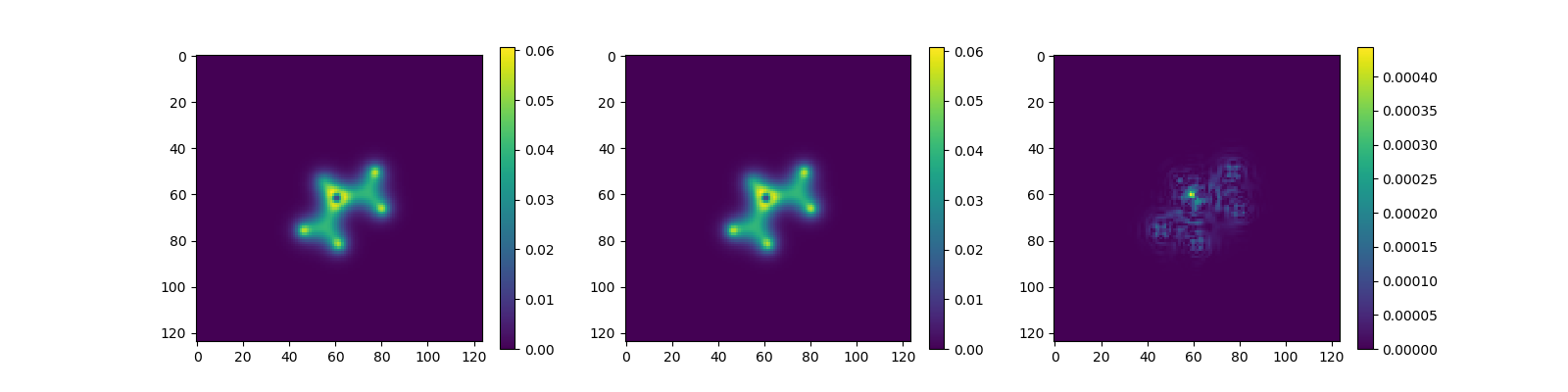}
     \includegraphics[width=\textwidth,clip]{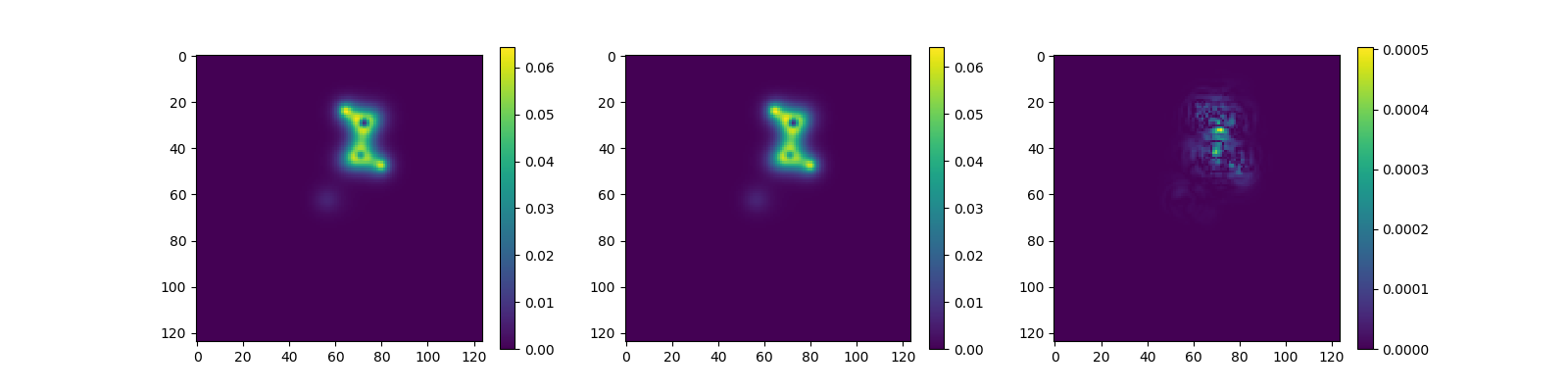}
     \caption{(left column) Slice of a snapshot of the electron density containing the largest point-wise error , (center column) slice of the density computed using the network, (right column) absolute error. Rows starting from the top : results for ethane, isobutane, and butane.} 
     \label{fig:C2H6}
 \end{figure}

Fig.~\ref{fig:C2H6} shows the slice with the largest error for a snapshot in the test set for both molecules. The relative test errors for the molecules are shown in Table \ref{tab:mol_summary}. The relative $\ell^1$ error in \cite{grisafi2018transferable} for ethane and butane are $1.14\%$ and $1.19\%$, respectively. Therefore our error is $6.1$ and $3.8$ times smaller, respectively. 
Finally Fig.~\ref{fig:water_all} (Appendix \ref{app:numer3D}) summarizes the distribution of the prediction error for the three molecules. We  observe that the distribution remains exceptionally close to the diagonal, thus indicating a very low error.

\begin{table}[ht] 
    \centering
    \begin{tabular}{| c |c  |c |}
        \hline
        Molecule \textbackslash error & $\texttt{err}_{\ell^2}$ & $\texttt{err}_{\ell^1}$ \\
        \hline 
        ethane  & $0.172$\% & $0.186$\% \\  
        isobutane & $0.194$\% & $0.222$\% \\  
        butane  & $0.289$\% & $0.314$\% \\ 
        \hline 
    \end{tabular}
    \caption{Error of the testing samples for different molecules. 
    }
\label{tab:mol_summary}
\end{table}

\subsubsection{H$_2$O}
 
\begin{figure}
     \centering
     \includegraphics[width=\textwidth,clip]{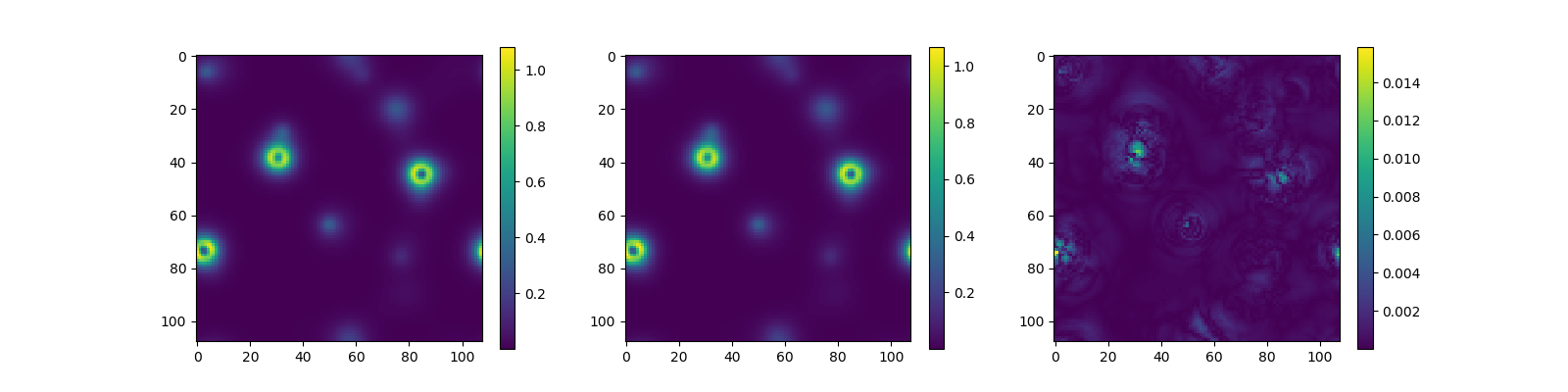}
     \caption{(left) slice of a snapshot of the electron density with $32$ water molecules, (center) slice of the density computed using the network, (right) slice containing the largest point-wise absolute error for the snapshot.}
     \label{fig:H20_32}
 \end{figure}

For water the kinetic energy cut-off is set to $40$ a.u.~
Both the training and test systems consist of $32$ water molecules. For the feature networks in Eq.~\eqref{eqn:hnetwork}, we use a dense linear network with four layers, each one containing  $\{5,10,20,40 \}$ nodes respectively. The ResNet fitting network uses $5$ dense layers and $50$ nodes per layer, where the skip connections are weighted by a trainable coefficient. The cutoff radius is $R_{c} = 3.5 \AA$. We trained the network a few times changing the random seed and we picked the one with the smallest generalization error. The error for a test snapshot with $32$ atoms is showcased in Fig.~\ref{fig:H20_32}, where we provide the slice containing the largest point-wise error.


\begin{table}[ht] 
    \centering
    \begin{tabular}{| c |c  |c |}
        \hline
        $N_{\texttt{mol}}$(H$_2$O)\textbackslash error & $\texttt{err}_{\ell^2}$ & $\texttt{err}_{\ell^1}$ \\
        \hline 
        $32$  & $0.606$\% & $1.080$\% \\  
        $64$  & $0.612$\% & $1.021$\% \\  
        $128$ & $0.503$\% & $0.891$\% \\  
        $256$ & $0.520$\% & $0.903$\% \\  
        $512$ & $0.528$\% & $0.921$\% \\
        \hline 
    \end{tabular}
    \caption{Error of the testing samples for different number of water molecules.}
\label{tab:H2O_summary}
\end{table}

Next we test the transferability of our model using systems with different sizes, which consists of $64$, $128$,  $256$ and $512$ water molecules, respectively. The largest system has a total number of 20,213,648 grid points. The relative $\ell^2$ and $\ell^1$ errors are summarized in Table~\ref{tab:H2O_summary}, where we can observe that inference error remains almost constant across different systems, which is around $0.5\%$ for the $\ell^2$ relative error (and $1.0\%$ for the $\ell^1$ relative error).
Finally Fig.~\ref{fig:water_all} (Appendix \ref{app:numer3D}) summarizes the distribution of the prediction error as we increase the system size. We can observe that the distribution remains very well concentrated within the diagonal thus indicating a very low error. 
Fig.~\ref{fig:H20_summary}  (Appendix \ref{app:numer3D}) shows the generalization error measured by the slice with the largest error for a snapshot in the test set. It demonstrates that the Deep Density network, learned using a small sized system, has excellent transferability to large systems.
The magnitude of the relative error agrees with that in Table~\ref{tab:H2O_summary}. In this case, we observe that the error is mostly concentrated on high-gradient portions of $\rho$ near the nuclei.

\subsubsection{Aluminum}
 
For the aluminum systems, we follow the same training pipeline as for the water case, and the kinetic energy cutoff is 20 a.u., and the largest system has total number of 1,906,624 grid points. The feature and fitting networks are chosen in a similar fashion to the water case. However, we used a large truncation radius $R_{c} = 6 \AA$. In this case, however, the initialization of constants in \eqref{eq:ansatz} was modified by using a larger truncation radius in order to start with a more uniform initial density .

The training stage was performed using a system with $32$ atoms as explained in Appendix \ref{section:parameters}. In Fig.~\ref{fig:Al_summary} (Appendix \ref{app:numer3D}) we demonstrate that the trained model is able to efficiently recover the peaks concentrated at the center of each nuclei, which accounts mainly for the electron density associated to  semi-core orbitals. Note that the ONCV pseudopotential treats all the 2s and 2p orbitals as semi-core electrons. As a result, the electron density in Al has sharp peaks and relatively large magnitudes. We test the transferability following the same procedure as for the water system. In Fig.~\ref{fig:Al_summary} we compare the electron density provided by the network and those from KS-DFT calculations for configurations containing $108$ and $256$ aluminum atoms. For both cases we observe a relative $\ell^2$ error below $2.5\%$ (see Table \ref{tab:Al_summary}). This is larger compared to that of the water system. From Fig.~\ref{fig:Al_summary} and Table \ref{tab:Al_summary} we observe that the errors also grow with respect to the system size. This may be due to the quality of the training data generated by PWDFT, which only uses the $\Gamma$-point sampling of the Brillouin zone, and the system size is relatively small. To verify this, we train the network with just $4$ snapshots of the $3\times 3 \times 3$ configuration. The relative test $\ell^2$ and $\ell^1$ error can be improved to $0.5\%$ and $1.2\%$, respectively, for a $3\times 3 \times 3$ configuration as shown in Fig.~\ref{fig:Al_3x3x3}. In addition, even though the absolute error of water and Al systems are comparable to each other, further inspection of Figs.~\ref{fig:H20_summary} and \ref{fig:Al_summary} reveals some qualitative difference between the two systems: the errors from the Al systems appear to be much more spatially delocalized, and hence it is possible that the errors are mainly contributed by the valence electrons rather than the semi-core electrons. To verify this, we tested our method with another data set, which uses the same configurations, but with the density generated by the Vienna \textit{ab initio} simulation package (VASP, version
5.4.4)~\cite{KresseFurthmuller1996,kresse1996efficient}, which do not treat electrons at 2s and 2p orbitals as semi-core electrons. The results are reported in Appendix \ref{app:numer3D}.

\section{Conclusion}\label{sec:conclude}

Leveraging the success of the recently developed Deep Potential, we propose the Deep Density method to use machine learning to bypass the solution of the Kohn-Sham equations, and to obtain the self-consistent electron density in the context of \textit{ab initio} molecular dynamics simulation. We demonstrate that the localization principle not only holds for insulating systems, but at least to some extent is also valid for metallic systems due to screening effects. Numerical results in one-dimensional systems and small molecular systems demonstrate that our construction can be very accurate, using a relatively small number of training samples. Our model can also be used to predict the electron density in the condensed phase, and can achieve excellent transferability for systems with up to $512$ water molecules.

In $1$D we have shown that this approach is able to efficiently compute the electron density for toy models emulating insulating, metallic and mixed system (to single precision). In $3$D deep density is able to learn the electron density for realistic chemical systems. However, we also observe that the accuracy of our model deteriorates when applied to real $3$D metallic systems such as aluminum. This may be caused by insufficient screening compared to toy models, but we also expect that our results may be further improved by employing different neural network architectures. 

We envisage to accelerate the current algorithm. The complexity for the point-wise evaluation of the electron density is linear with respect to the systems size. However, a simple modification of the proposed approach can lead to a point-wise time evaluation that is independent of the systems size, thus producing a linear scaling algorithm. Another line of work is to improve the efficiency of training and prediction. In the current implementation, the descriptors are computed on CPUs, and this becomes a bottleneck when the electron density 
on millions of data points or more need to be evaluated. We expect that by employing a GPU based implementation and by computing the electron density at different grid points in an embarrassingly parallel fashion, the efficiency can be greatly improved.
We expect that these improvements would make Deep Density to be a very useful tool for the analysis and prediction of electronic structures.


\section*{Acknowledgment}

This work was partially supported by the Department of Energy under Grant No. DE-SC0017867 and the CAMERA program (L.~L., L.~Z.-N., J.~Z.), as well as  the Center Chemistry in Solution and at Interfaces (CSI) funded by the DOE Award DE-SC001934 (Y.~C., L.~Z.).  We thank the Berkeley Research Computing (BRC) program at the University of California, Berkeley, 
the TIGRESS High Performance Computer Center at Princeton University,
the National Energy Research Scientific Computing Center (NERSC), and the Google Cloud Platform (GCP) for the computational resources.  We thank Roberto Car, Weinan E, Yuwei Fan, Jiequn Han, Yu-hang Tang, Han Wang, Chao Yang, Lexing Ying for valuable discussions at various stages of the project.

\bibliography{mldensity,ref2}
\bibliographystyle{ieeetr}

\appendix

\section{Numerical results for 1D systems}\label{app:numer1D}

We use a $1$D reduced Hartree-Fock model similar to the one presented in~\cite{LinYang2013}. This simplified model depends nonlinearly on the electron density $\rho$ only through the Hartree interaction, and does not include the exchange-correlation functional. However, it can still qualitatively reproduce certain phenomena in $3$D, such as the difference between insulating and metallic systems, and the screening effect shown in Section~\ref{section:preliminaries}. The Hamiltonian in our $1$D model is given by (we still use the notations $\vr$ and $\vR$ though this is a one-dimensional system)
\begin{align}
    H[\rho, \{\vR_I\}_{I =1}^{N_a}] & = -\frac{1}{2}\frac{d^2}{d\vr^2} + V_{\text{hxc}}[\vr; \rho] + V_{\text{ion}}[\vr;\{\vR_I\}_{I =1}^{N_a}], \\
                                    & = -\frac{1}{2}\frac{d^2}{d\vr^2} + \int K(\vr, \vr')(\rho(\vr')+m(\vr'; \{\vR_I\}_{I =1}^{N_a})) \ud\vr'.
\end{align}
Here we use a pseudopotential to represent the electron-ion interaction, and the total pseudo charge density is given by 
\begin{equation}
    m(\vr; \{\vR_I\}_{I =1}^{N_a}) = \sum_{I=1}^{N_a} -\frac{Z_I}{\sqrt{2\pi \sigma_I^2}}\exp{\left(-\frac{1}{2\sigma_I^2}(\vr-\vR_I)^2\right)}.
\end{equation}
Here $Z_I$ represents the charge of $I$-th nucleus, and $\sigma_I$ represents the width of the nuclei potential within the pseudopotential theory. $\sigma_I$ is tuned so that $I$-th nucleus can qualitative behave as a metal or as an insulator.
Since the standard Coulomb interaction diverges in $1$D, we employ the Yukawa kernel
\begin{equation}
    K(\vr, \vr') = \frac{2\pi}{\kappa \epsilon_0} e^{\kappa |\vr-\vr'|},
\end{equation}
where the parameters $\kappa=0.01$ and $\epsilon_0=10.0$ are fixed constants throughout the experiments. Our units here are arbitrary, but will be referred to as the atomic unit (a.u.) for simplicity.

The Kohn-Sham equations are solved using the standard self-consistent field iteration \cite{Martin2008} method. In particular, we use Anderson mixing~\cite{Anderson:1965:IPN:321296.321305} of the potential with mix dimension $10$.

\begin{figure}[htbp]
{%
\subfigure[]{%
\label{fig:inseig}
\includegraphics[width=0.3\textwidth]{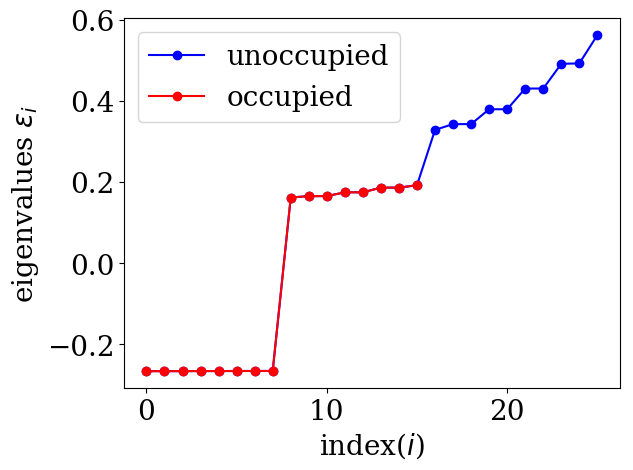}
} 
\subfigure[]{%
\label{fig:meteig}
\includegraphics[width=0.3\textwidth]{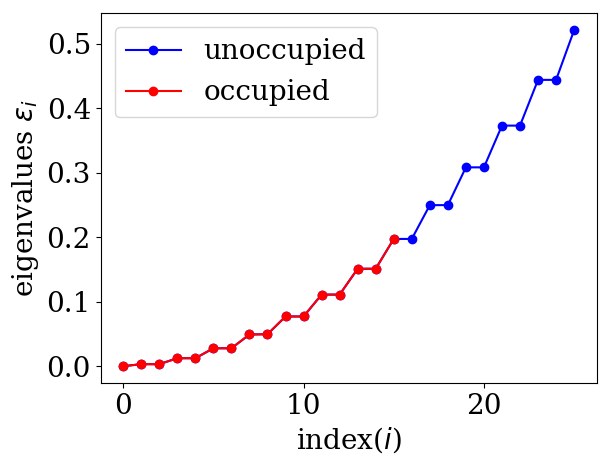}
}
\subfigure[]{%
\label{fig:halfnhalfeig}
\includegraphics[width=0.3\textwidth]{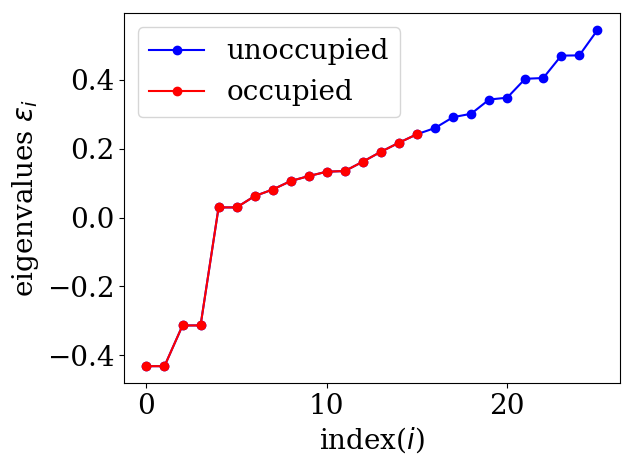}
}
\caption{Eigenvalues for (a) the insulating system, $\sigma_I=1.0$ for all $I$; (b) the metallic system, $\sigma=6.0$ for all $I$; (c) the mixed metallic-insulating system, $\sigma_I=1.0$ for $1\leq I \leq 4$ and $\sigma_I=6.0$ for $5 \leq I \leq 8$.} 
\label{fig:1Deigenvalues}
}
\end{figure}

We study three types of systems: the insulating system, the metallic system, and the mixed metallic-insulating system. 
Fig.~\ref{fig:1Deigenvalues} displays the occupied eigenvalues and the first ten unoccupied eigenvalues for all the three systems. 
In particular, when the system is metallic or has a mixed metallic-insulating character, the self-consistent field iteration can be very difficult to converge due to the small energy gaps and charge sloshing behavior \cite{LinYang2013}.

We consider a periodic system that includes 8 atoms in the unit cell, with 2 electrons per site, i.e.  $Z_I=2$ for $I=1, ..., 8$. The atoms are $10$ a.u. apart, located at $5, 15, 25, \cdots, 75$. The size of the supercell in this case is thus 80 a.u.
As mentioned above, by adjusting $\sigma_I$ we can obtain different qualitative behaviors. 
On the one hand, when $\sigma_I = 1.0$, the model qualitatively behaves as an insulating system with an energy gap of $0.136$.
On the other hand, if $\sigma_I = 6.0$, then the model qualitatively behaves as a metallic system and its energy gap is  $5.5\times 10^{-8}$ (i.e. the system is gapless).
To test the ability of the proposed architecture to deal with interactions between atoms of different species, we also introduce a mixed metallic-insulating system, which is obtained by setting $\sigma_I=1.0$ for $I=1,2,3,4$, and $\sigma_I=6.0$ for $I=5,6,7,8$. The energy gap in this case equals to $0.018$.
Fig.~\ref{fig:1Deigenvalues} displays the occupied eigenvalues and the first ten unoccupied eigenvalues for all the three systems. 

\begin{table}
\centering
 \begin{tabular}{|c c c c c c c c|} 
 \hline
 $N_{\text{near}}$ & 1 & 2 & 3 & 4 & 5 & 6 & 7 \\ [0.5ex] 
 \hline\hline
 Insulator & 1.91E-08 & 4.02E-08 & 2.26E-08 & 5.02E-08 & 1.38E-07 & 1.39E-07 & 1.13E-07 \\ 
 \hline
 Metal & 9.11E-10 & 1.08E-09 & 1.04E-09 & 1.10E-09 & 3.00E-09 & 7.65E-09 & 3.53E-09 \\
 \hline
\end{tabular}
\caption{Validation error (MSE) for single type system with different values of $N_{\text{near}}$ }
\label{table:singletype}
\end{table}

For the 1D problem, our implementation is purely based on \texttt{python} and \texttt{tensorflow}. 
Furthermore, 1D model does not involve the rotational degrees of freedom. This allows us to simplify the network structure in Section \ref{section:architecture} as below.

For simplicity of implementation, we introduce a parameter $N_{\text{near}}$ instead of cutoff $R_c$, so the index set $\cI_{N_{\text{near}}}(I)$ is decided by choosing the indices of the nearest $N_{\text{near}}$ atoms. 
The model is constructed using the same ansatz as Eq.~\eqref{eq:ansatz} where $\cN^I$ and $ \cE^I$ are neural networks. However, the construction of the input to these networks, which are the descriptors $\cD^I (\vr, \cR^I)$ defined in Section \ref{section:architecture} as Eq.~\eqref{eqn:dnetwork}, is much simpler. To define the descriptors, we start with $\vd^I_J = [R_{JI}, \frac{1}{R_{JI}}(\vR_J-\vR_I)]$ for $J \in \cI_{N_{\text{near}}}(I)$ (distance information and direction information). Since we follow the form in Eq.~\eqref{eq:dIJ} with $\vR_J-\vR_I$ reduced to one dimension, each $\vd^I_J$ is in $\mathbb{R}^2$. 
The electron information $\vd^I_0 \in \mathbb{R}^2$ is fed to the descriptor directly, whereas the atom information $\vd^I_J, J\neq 0$, is passed to the function $g_{s(I),s(J)}$ before being fed to the descriptor. 
For the insulating system and the metallic system, we only have one such function $g$, whereas for the mixed metallic-insulating system, we have four $g_{s(I),s(J)}$ networks because each of $s(I),s(J)$ can be one of the two types. Given that rotation symmetry in 1D is trivial, the descriptor $\cD^I$ is formed simply by concatenating the electron information, $\vd^I_0$, and atom information, $g_{s(I),s(J)}(\vd^I_J)$. 
The output of $g$ is of $M$ dimension and there are $N_{\text{near}}$ number of nearby atoms, so descriptors are $\cD^I \in \mathbb{R}^{2 + MN_{\text{near}}}$. 

To treat the mixed metal-insulator system with two types of atoms, we implement a control flow so that at run time, the model knows which $g_{s(I),s(J)}$ to apply based on species of atom $I$ and $J$. 
For simplicity we incorporate the information of the species as follows. Let range of $s(J)$ be $\{1,2\}$. We encode the two atom types as vectors $\vv_1=[1,0]^T, \vv_2=[0,1]^T$. For a fixed center atom $I$, and adjacent atom $J$, we pass $\vd^I_J$ to $g_{s(I),s}$ for both $s=1,2$ and then calculate the output 
$$
g_{s(I),s(J)}(\vd^I_J) = \left(\vv_{s(J)}^T\vv_1\right)g_{s(I),1}(\vd^I_J)+
\left(\vv_{s(J)}^T\vv_2\right)g_{s(I),2}(\vd^I_J).
$$

The training and test data sets are generated through molecular dynamics simulations. We use the Verlet algorithm \cite{FrenkelSmit2002} for the time propagation, where the forces are computed using the Hellmann-Feynman formula.
At each time step we store the atomic configuration, $\{\vR_I\}_{I = 1}^8$, and the corresponding self-converged electron density, $\rho$, generated from the KS-DFT computation. For all three systems, we use the first $8000$ snapshots for training and the next $400$ snapshots for validation. In order to reduce the correlation of the shots and the amount of training time, we down-sample the training snapshots by a factor $80$, i.e., we take 100 evenly time-spaced snapshots. The same procedure is applied to the validation snapshots. We then use these $100$ training snapshots and $5$ validation snapshots to train the network. For the mixed metallic-insulating system, the number of trainable parameters increases because of the four $g_{s(I),s(J)}$ networks, so we reduce the down-sampling factor to have more training snapshots (e.g. down sample the first 8000 snapshots by a factor of 20 to obtain 400 training snapshots). We also find that if we only use $100$ training snapshots, the relative $\ell^1$ error can increase and be higher than $1\%$. This indicates that the mixed insulating-metallic system is indeed more difficult and requires a larger number of training samples.

The training is performed using standard Adam optimizer \cite{kingma2015adam} and a mean squared error loss. Given the simplicity and small scale of the problem we visit all the points at each snapshot, in contrast with the 3D training that will require importance sampling for efficiency consideration. The network was trained for 400 epochs, the model with lowest validation loss was saved. For each hyperparameter setting (Fixed $N_{\text{layer}}$, $N_{\text{nodes}}$, $N_{\text{near}}$), we run 5 experiments and report the one with lowest validation error. The validation errors are measured using mean squared error (MSE), namely \begin{equation}
    \frac{1}{N_{\text{validation}}} \sum_{j=1}^{N_{\text{validation}}}\left(\varrho(\vr_{j}, \{R_I\}) - \varrho_{NN}(\vr_{j}, \{R_I\}\right)^2.
\end{equation}

In Table \ref{table:singletype}, we  observe that the validation loss reaches well below 1E-06. Another observation is that the network is relatively insensitive to the hyperparameter $N_{\text{near}}$ here, even when the system is gapless. Thus we  fix $N_{\text{near}}=2$ for the mixed metallic-insulating system model for simplicity. 

\begin{table}
\centering
 \begin{tabular}{|c c c c|} 
 \hline
 $N_{\text{sample}}$ & $100$ & $200$ & $400$\\ [0.5ex] 
 \hline\hline
 Two-atom-type & 1.87E-07 & 6.94E-08 & 4.40E-08 \\ 
 \hline
\end{tabular}
\caption{Validation error (MSE) for two-atom-type system with increasing training samples}
\label{table:doubletype}
\end{table}
In Table \ref{table:doubletype}, $N_{\text{sample}}$  is the number of snapshots in training, so the real amount of training data is the number of snapshots multiplied by the number of grid points for the $1$D electron density. The validation loss reaches below 1E-07 once we increase the number of snapshots to $200$.

\section{Numerical results for 3D systems}\label{app:numer3D}

\subsection{Simulation parameters} \label{section:parameters}

The parameters in the ansatz are initialized after a precomputation step that depends on each setup. This precomputation involves the following steps: selecting one atom for each species, sampling the density within a small radius of that atom, and computing the parameters $A_{s(I)}, B_{s(I)}, C_{s(I)}$, and $D_{s(I)}$ that best fits the sampled density, without the neural networks, using standard quasi-Newton optimization methods. The purpose of this precomputation step is to help the optimization find a suitable minimum. The weights in the Neural Network are initialized using a normalized Gaussian distribution. The objective function is the mean squared loss. 

For the training stage we use the Nadam optimizer \cite{kingma2015adam} with an exponential scheduling, in which for every $20000$ iterations we decrease the learning rate by a factor $0.95$, and we initialize the learning rate as $0.003$. At each iteration we draw $n_{\texttt{s}}=64$ samples from the snapshots. The training is scheduled as follows: we train the network for a million iterations using only $5$ snapshots, then we train the network for another million iterations using $20$ snapshots and finally we train the network for two million iteration using $80$ snapshots. The remaining $20$ snapshots were used for testing throughout the training.

\begin{figure}
     \centering
     \includegraphics[width=\textwidth,clip]{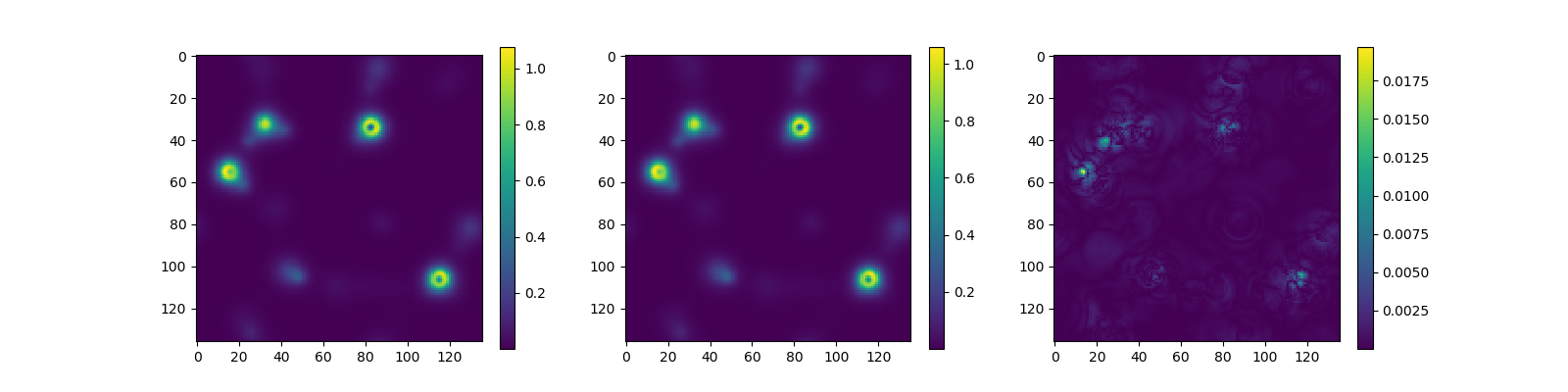}
     \includegraphics[width=\textwidth,clip]{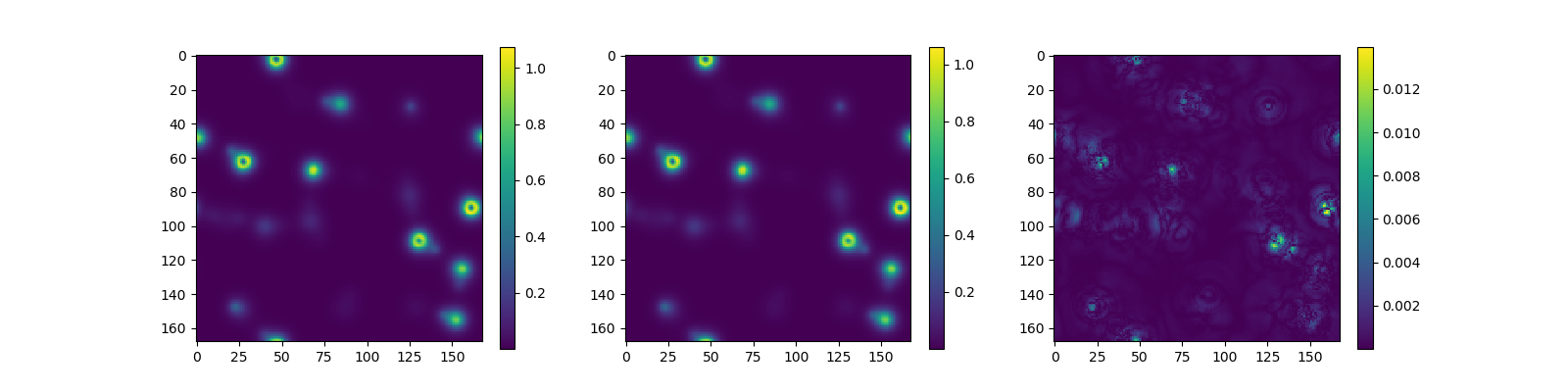}
     \includegraphics[width=\textwidth,clip]{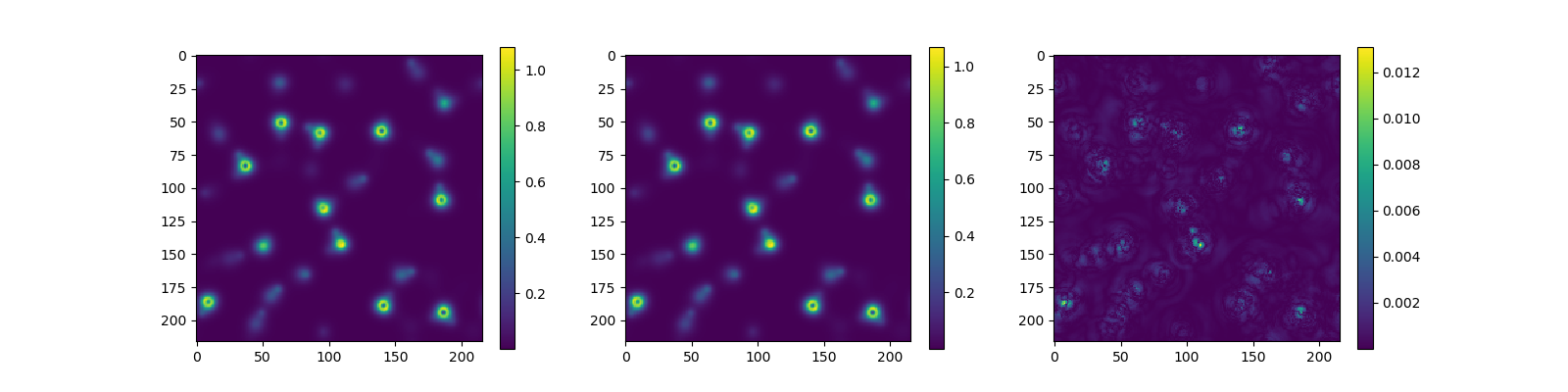}
     \includegraphics[width=\textwidth,clip]{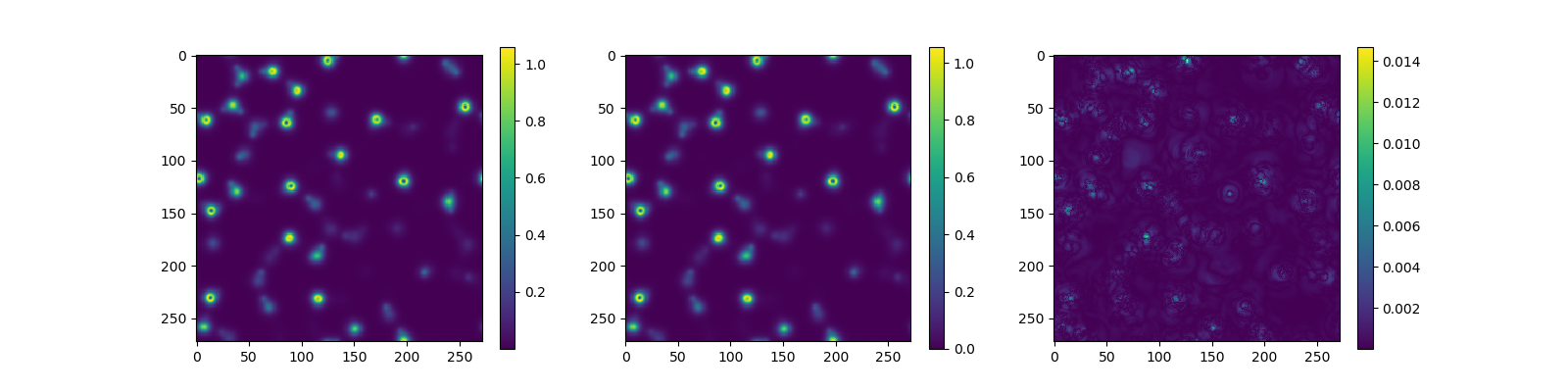}     
     \caption{(left column) slice of a snapshot of the electron density, (center column) slice of the density computed using the network, (right column) slice of the absolute error with the highest point-wise absolute error.  rows starting from the top : results for the system containing $64$, $128$, $256$ and $512$ water molecules }
     \label{fig:H20_summary}
 \end{figure}

At each iteration $n_{\texttt{s}}$ samples are extracted from the training data. Each sample represents a pixel of the images shown in Fig.~\ref{fig:H20_summary}. For the water system we have $80$ training snapshots and each snapshot contain around $1.24$E$6$ pixels, totalling roughly $100$ million data points. For the aluminum system we have the same amount of training snapshots but each snapshot contain around $2.564$E$5$ pixels, totalling roughly $21$ million data points. For the small organic molecules we have roughly $125$ million data point for each system. Thus we need to visit them judiciously in order to be efficient. Given that different systems may have very different localization properties we use a sampling strategy based on the norm of the density at each pixel. In particular, during the training stage the samples are drawn following the distribution $|\rho(\vr)|^{\alpha}$, where the value of $\alpha$ is tuned for each setup, in order to avoid visiting small values of the densities too often, thus improving the efficiency of training. In particular we used $1/2$ for the small organic molecules, $6/5$ for the water systems, and $1$ for the aluminum systems.

We estimate the test error by comparing the result given by the network against the test snapshots in the small system, and we estimate the transferability of the algorithm by comparing the electron density generated by the trained model for the larger systems versus the one computed using PWDFT. 

The computation of the electron density were performed at the NERSC cluster Cori, which is comprised of $2,388$ dual socket nodes with $32$ cores and $256$ GB of RAM, whereas, the training of the models and inference steps were performed in a 16 core machine used with 64 GB of RAM and a Tesla V100 GPU with 16GB memory.

\subsection{Additional plots of the organic molecules and water systems}
\begin{figure}[htp]
    \centering
    \includegraphics[width=0.45\textwidth,clip]{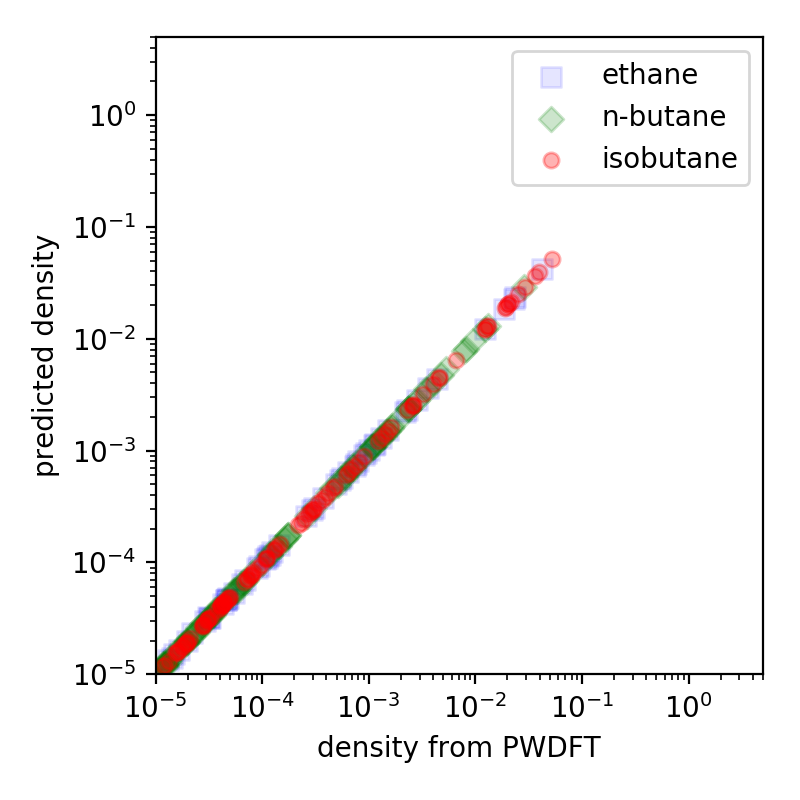}
    \includegraphics[width=0.45\textwidth,clip]{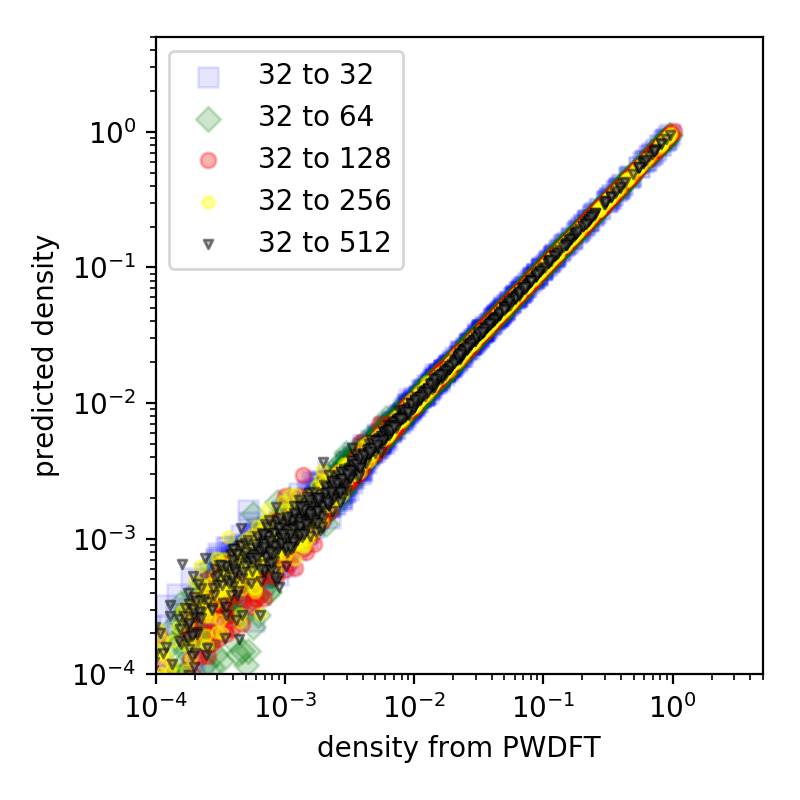}
     \caption{(left) scatter plot of the predicted and test densities for the different small organic molecules, (right) scatter plot of the predicted and test densities for different water systems.}
     \label{fig:water_all}
 \end{figure}

In addition to the figures in the main text, we include Fig.~\ref{fig:water_all}, which depicts the performance of Deep Density for the different small molecules and different water systems. Fig.~\ref{fig:water_all} represent a scatter plot in logarithmic scale of the value of the predicted density and the density computed with PWDFT for the same configuration and sampling points. We can observe that for the former the error are almost negligible, and the later the errors are higher but are still very small. 

\subsection{Additional results for aluminum}

Our VASP calculation also used the PBE exchange-correlation functional, but with the projector augmented wave method (PAW) \cite{Blochl1994} to handle the core electrons. In particular, the 2s and 2p orbitals are treated as core electrons, and hence there are only $3$ valence electrons per atom. The kinetic energy cutoff for the plane wave expansion is set to $600$ eV, and the Brillouin zone is sampled with the Monkhorst-Pack mesh~\cite{MonkhorstPack1976} at the spacing $h_k$ = 0.08 $\AA^{-1}$.  
The order 1 Methfessel-Paxton smearing method with $\sigma = 2900$ K is adopted.
The self-consistent field (SCF) iteration stops when the total
energy and band structure energy differences between two consecutive
steps are smaller than $10^{-6}$ eV.
In this case the density contains only the contribution of valence electrons. 

\begin{figure}[ht]
    \centering
    \includegraphics[width=0.45\textwidth,clip]{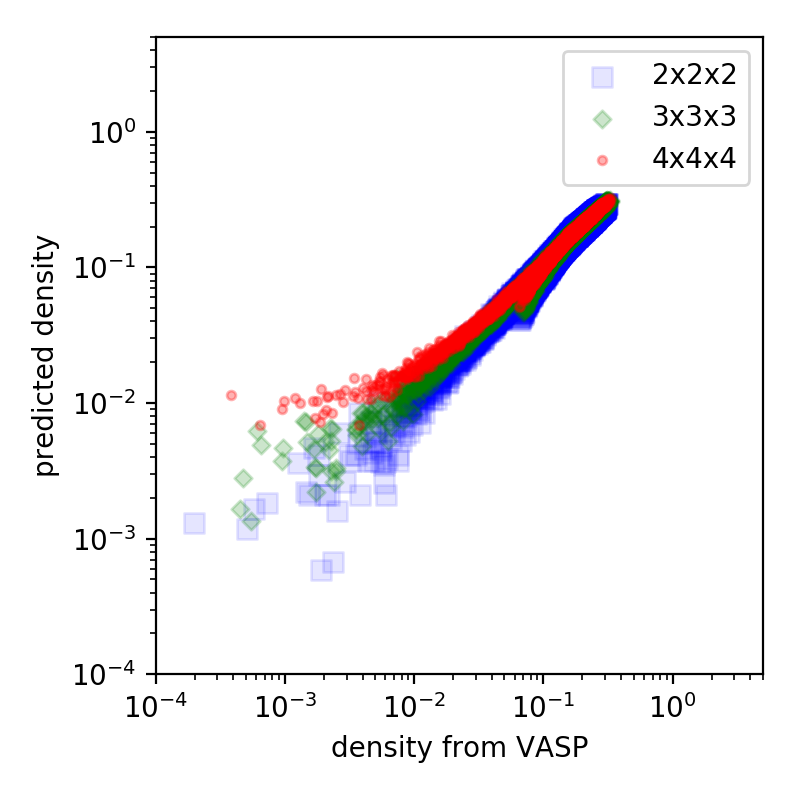}
    \includegraphics[width=0.45\textwidth,clip]{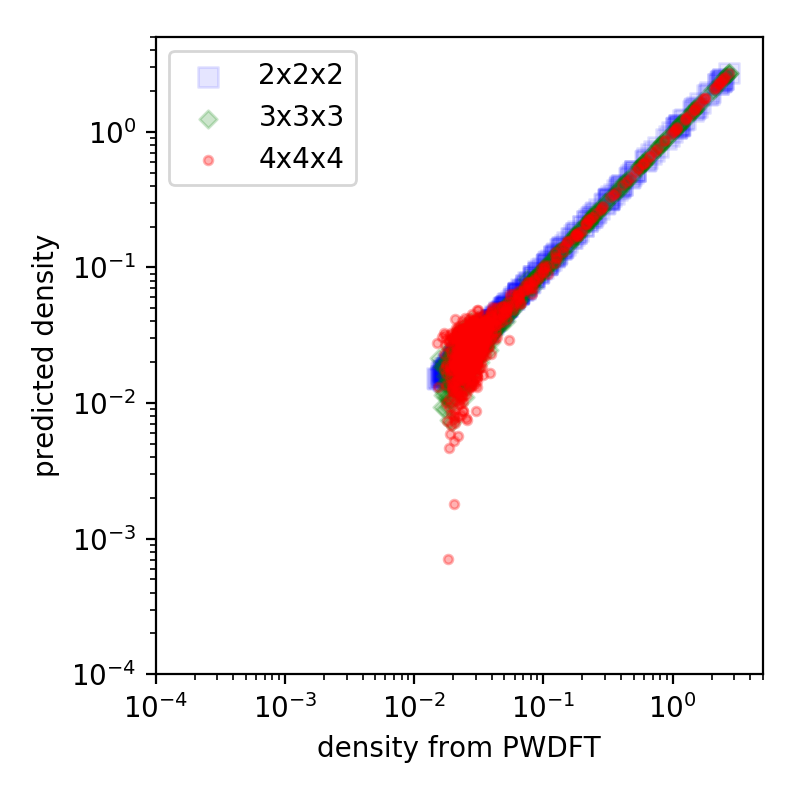}
     \caption{(left) scatter plot of the predicted and test densities generated by VASP for the different aluminum systems, (right) scatter plot of the predicted and test densities generated by PWDFT for the different aluminum systems. The magnitudes of the density from PWDFT are higher due to the inclusion of semi-core electrons.}
     \label{fig:al_all}
 \end{figure}

\begin{figure}
     \centering
    \includegraphics[width=\textwidth,clip]{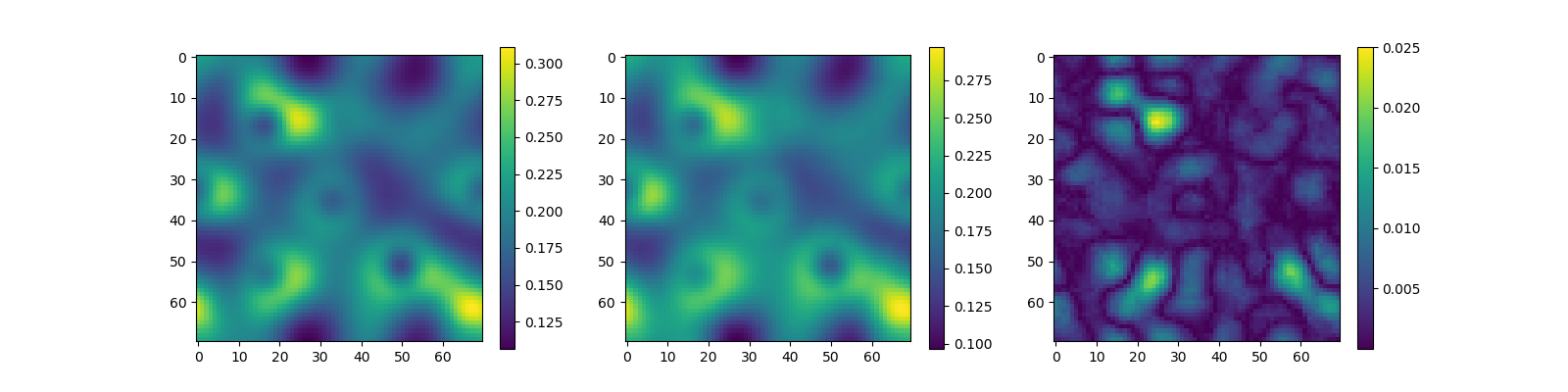}
     \includegraphics[width=\textwidth,clip]{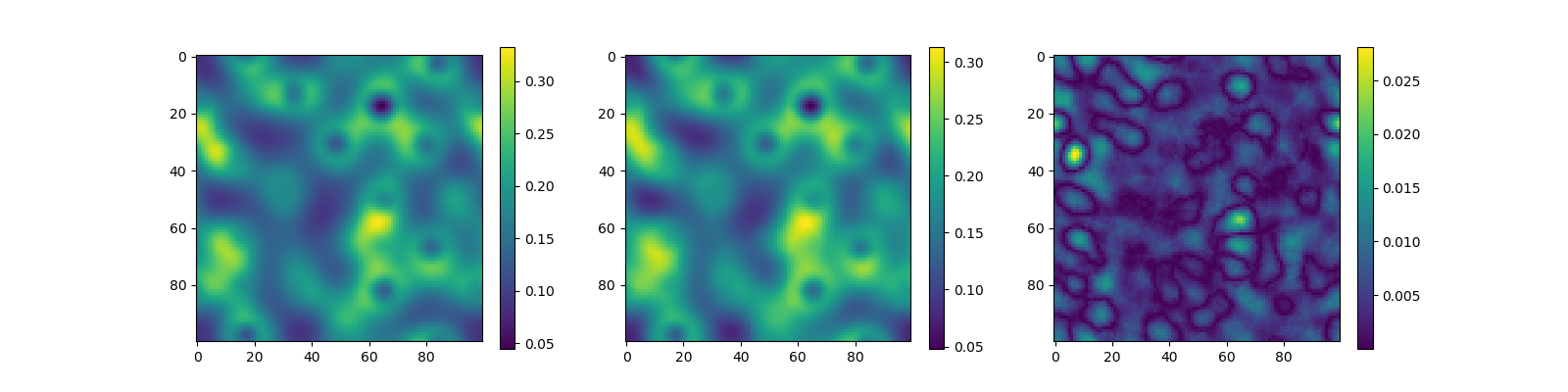}    
     \includegraphics[width=\textwidth,clip]{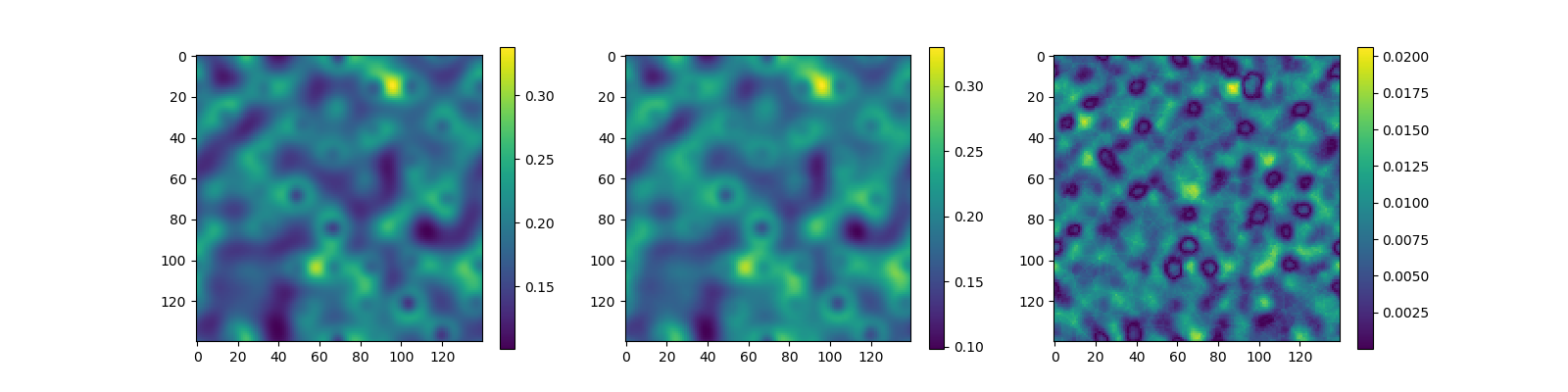}
     \caption{(left column) slice of a snapshot of the electron density from VASP, (center column) slice of the density computed using the network, (right column) slice of absolute error containing the largest point-wise error. Rows starting from the top : results for the system containing $32$, $108$, and $256$ aluminum atoms following $2\times2\times2$, $ 3\times3\times3$, and  $4\times4\times4$ configurations respectively.}
     \label{fig:Al_VASP_summary}
 \end{figure}

We used the same training pipeline as before. We perform training using a number of snapshots for a system containing $2\times2\times2$ unit cells, and test the network for a number of systems with $2\times2\times2$, $3\times3\times3$ and $4\times4\times4$ unit cells, respectively. 
The scatter plot in Fig. \ref{fig:al_all} suggests that the test error for the aluminum system is indeed much larger. Fig.~\ref{fig:Al_VASP_summary} shows the test error using the density generated with VASP. The error is largely delocalized, which confirms the previous study with PWDFT that most error originates from valence electrons.
In addition, from Table \ref{tab:Al_summary}, we observe that the generalization error still grows, albeit slightly slower, with respect to the system size, thanks to the refined Brillouin zone sampling when generating the training data set.

\begin{table}[ht] 
    \centering
    \begin{tabular}{| c |c  |c |c |c |}
        \hline
        $N_a$ (Al)$\textbackslash$ error & $\texttt{err}_{\ell^2}$ & $\texttt{err}_{\ell^1}$ & $\texttt{err}_{\ell^2}$ & $\texttt{err}_{\ell^1}$ \\
& PWDFT & PWDFT & VASP & VASP \\
        \hline 
        $32$  & $0.504$\% & $1.400$\% & $2.614$\% & $2.015$\%\\  
        $108$ & $1.512$\% & $3.937$\% & $3.040$\% & $2.529$\%\\  
        $256$ & $2.244$\% & $5.801$\% & $4.847$\% & $4.515$\%\\ 
        \hline 
    \end{tabular}
    \caption{Error of the testing samples for different number of atoms for Al. The data are generated using PWDFT (with semicore electrons) and VASP (without semicore electrons) respectively.}
\label{tab:Al_summary}
\end{table}

 \begin{figure}
     \centering

      \includegraphics[width=\textwidth,clip]{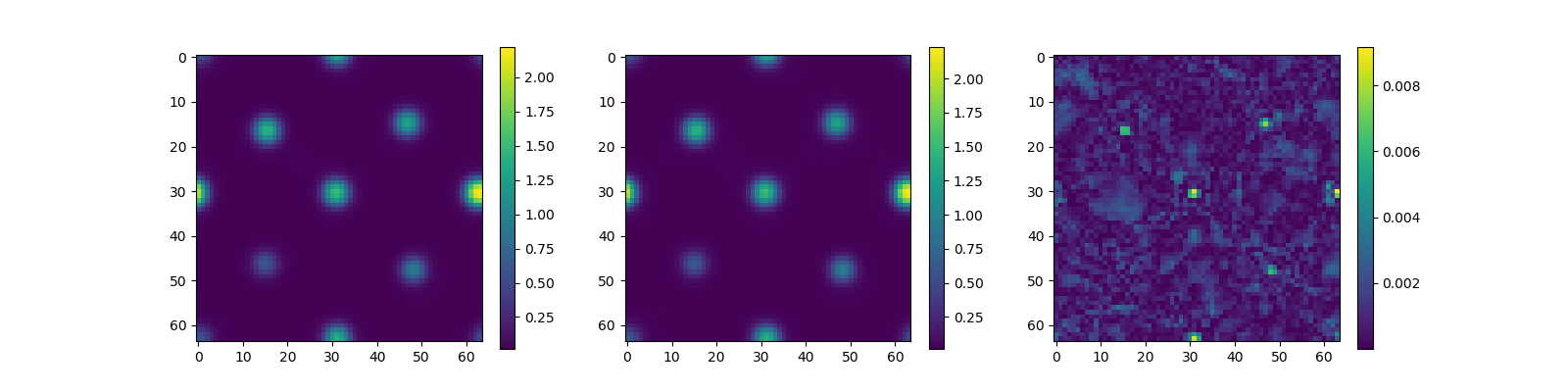}
     \includegraphics[width=\textwidth,clip]{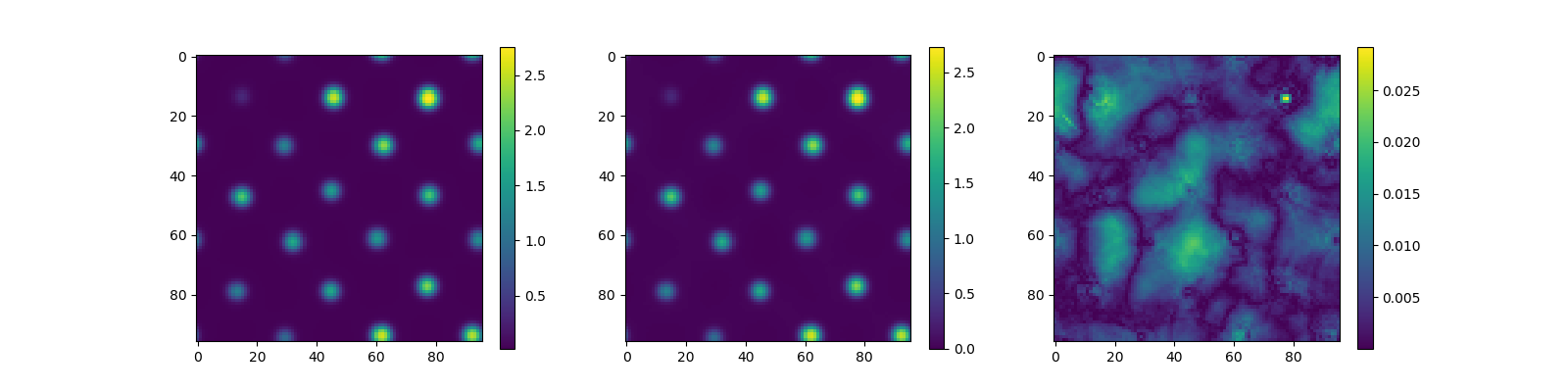}
     \includegraphics[width=\textwidth,clip]{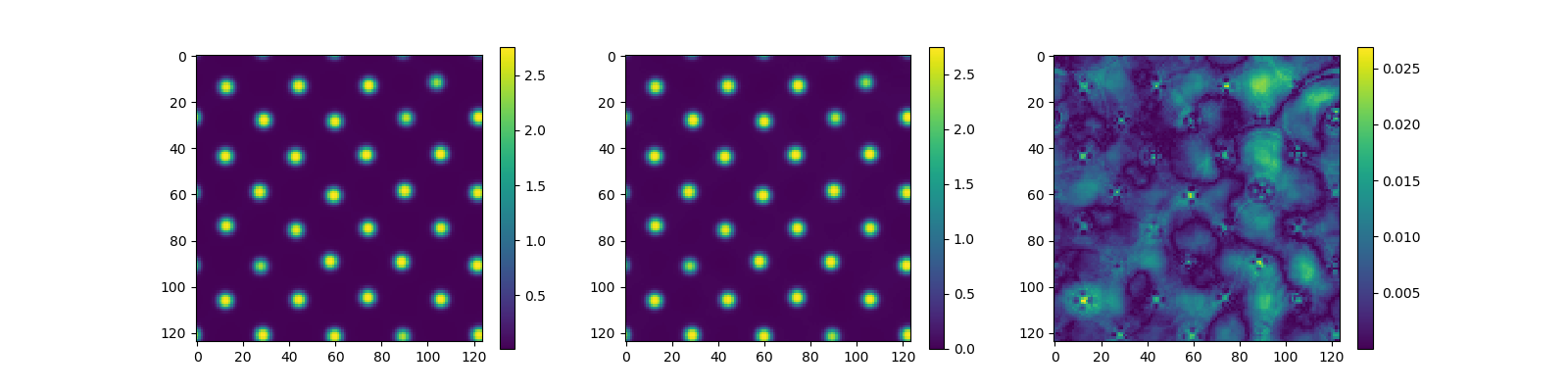}    
     \caption{(left column) slice of a snapshot of the electron density, (center column) slice of the density computed using the network, (right column) slice of the absolute error with the highest point-wise absolute error. Rows starting from the top : results for the system containing $32$, $108$, and $256$ aluminum atoms following $2\times2\times2$, $ 3\times3\times3$, and  $4\times4\times4$ configurations respectively. The calculations are performed using PWDFT.}
     \label{fig:Al_summary}
 \end{figure}

 \begin{figure}
     \centering

     \includegraphics[width=\textwidth,clip]{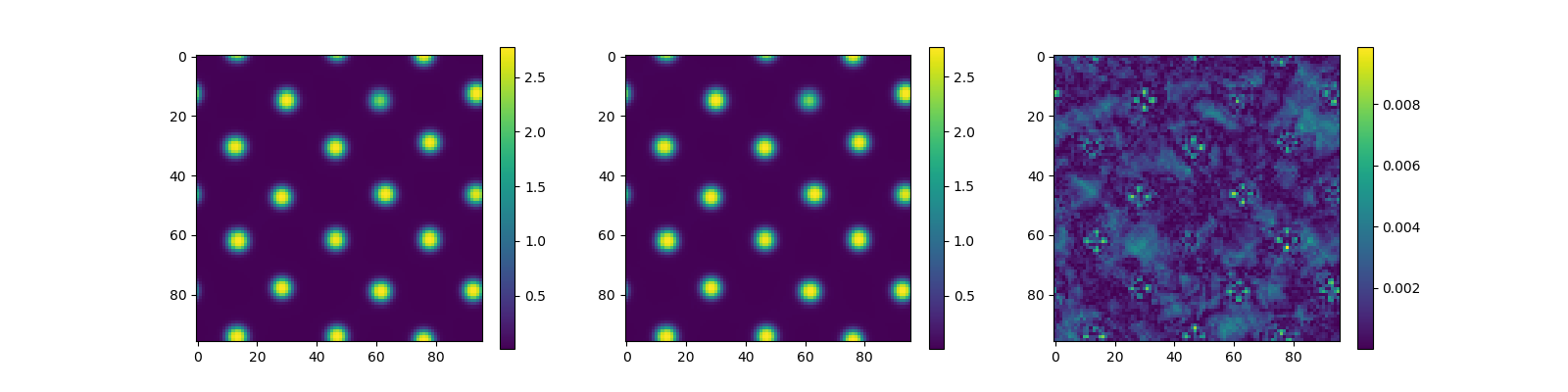}
     \caption{(left) slice of the snapshot produced by computing the electron density of $108$ aluminum atoms in a $3\times3\times3$ configuration, (center) slice of the density computed using the network which was re-trained using $4$ snapshots of the $3\times3\times3$ configuration, (right) slice of the absolute error with the highest point-wise absolute error. The calculations are performed using PWDFT.}
     \label{fig:Al_3x3x3}
 \end{figure}

\end{document}